\documentstyle[12pt]{article}
\makeatletter
\newtoks\@stequation
\def\subequations{\refstepcounter{equation}%
  \edef\@savedequation{\the\c@equation}%
  \@stequation=\expandafter{\theequation}%   %only want \theequation
  \edef\@savedtheequation{\the\@stequation}% %expanded once
  \edef\oldtheequation{\theequation}%
  \setcounter{equation}{0}%
  \def\theequation{\oldtheequation\alph{equation}}}
\def\endsubequations{%
  \ifnum\c@equation < 2 \@warning{Only \the\c@equation\space subequation
    used in equation \@savedequation}\fi
  \setcounter{equation}{\@savedequation}%
  \@stequation=\expandafter{\@savedtheequation}%
  \edef\theequation{\the\@stequation}%
  \global\@ignoretrue}

\def\eqnarray{\stepcounter{equation}\let\@currentlabel\theequation
\global\@eqnswtrue\m@th
\global\@eqcnt\z@\tabskip\@centering\let\\\@eqncr
$$\halign to\displaywidth\bgroup\@eqnsel\hskip\@centering
     $\displaystyle\tabskip\z@{##}$&\global\@eqcnt\@ne
      \hfil$\;{##}\;$\hfil
     &\global\@eqcnt\tw@ $\displaystyle\tabskip\z@{##}$\hfil
   \tabskip\@centering&\llap{##}\tabskip\z@\cr}
\makeatother

\renewcommand{\theequation}{\thesection.\arabic{equation}}

\newcommand{\VEV}[1]{{\langle {#1} \rangle}}
\def\beq{\begin{equation}}
\def\eeq{\end{equation}}
\def\bea{\begin{eqnarray}}
\def\eea{\end{eqnarray}}
\def\bseq{\begin{subequations}}
\def\eseq{\end{subequations}}
\def\nn{\nonumber}
\def\dfrac{\displaystyle\frac}
\def\numt#1#2{#1 \times 10^{#2}}
\def\degree#1{#1^{\circ}}

\def\ie{{\it i.e.,~}}
\def\eg{{\it e.g.~}}

\def\id{{\mit I}}

\def\PR#1#2#3{Phys. Rev. {\bf #1}, #2 (#3)}
\def\PRL#1#2#3{Phys. Rev. Lett. {\bf #1}, #2 (#3)}
\def\PL#1#2#3{Phys. Lett. {\bf #1}, #2 (#3)}

\def\NP#1#2#3{Nucl. Phys. {\bf #1}, #2 (#3)}

\def\PTP#1#2#3{Prog. Theor. Phys. {\bf #1}, #2 (#3)}

\def\eqref#1{eq.(\ref{eqn:#1})}

\def\eqlab#1{\label{eqn:#1}}

\def\Tbref#1{Table \ref{tbl:#1}}
\def\tblab#1{\label{tbl:#1}}

\def\bmaT{\left(\begin{array}{ccc}}
\def\emaT{\end{array}\right)}
\def\bma{\left( \begin{array} }
\def\ema{\end{array} \right)}

\def\l{\left}
\def\r{\right}

\def\Su{\rm u}
\def\Sd{\rm d}

\def\Se{\rm e}

\def\tr{{\rm tr}}

\def\hu{{y^{\rm u}}}
\def\hd{{y^{\rm d}}}
\def\he{{y^{\rm e}}}
\def\hn{{y^{\nu}}}

\begin{document}
\title{Quark and Lepton Flavor Mixings \\
in the SU(5) Grand Unification Theory}
\author{{K.~Hagiwara}
        ~~~~and~~~~
        {N.~Okamura}\thanks{E-mail address: naotoshi.okamura@kek.jp}\\
\\
{\it Theory Group, KEK, Tsukuba, Ibaraki 305-0801, Japan}}
\date{ }
\maketitle
\vspace{-8.5cm}
\begin{flushright}
hep-ph/9811495 \\
KEK-TH/605\\
Nov. 1998.
\end{flushright}
\vspace{5.7cm}
\begin{abstract}
 We explain the imbalance of the flavor mixing angles
between the quark and the lepton sectors
in the context of the SU(5) GUT with the see-saw mechanism.
 The quark masses and the CKM matrix elements are obtained
by using, respectively, the Fritzsch and Branco$-$Silva-Marcos form
for the up- and down-quark Yukawa matrices ($\hu$ and $\hd$)
at the GUT scale.
 The charged-lepton Yukawa matrix ($\he$) is the transpose
of $\hd$, modified by the Georgi-Jarlskog factor.
 We show that the neutrino masses and mixing angles suggested
by the recent solar and atmospheric neutrinos are
then obtained from a simple texture of the neutrino
Yukawa matrix ($\hn$) and a diagonal right-handed Majorana
mass matrix at the GUT scale.
\end{abstract}

\section{Introduction}
\label{sec:1}
\setcounter{equation}{0}

 The Standard Model (SM) is a successful theory.
 Current high energy experiments are explained within the SM.
 However, the SM cannot predict fermion masses and their flavor mixing.
 It is generally expected that there is a more fundamental theory
which gives the SM as its low-energy effective theory.
 The grand unified theory (GUT) is among the candidates
of a more fundamental theory.
 It is also believed that fermion masses and flavor mixings
are the keys to open the door to new physics beyond the SM.

 Recent neutrino experiments 
\cite{CHOOZ}\nocite{solar,solar2,solar3,solar4,Atm,Atm2,Atm3}-\cite{SK1}
have provided evidences that there are neutrino masses and
their flavor mixings.
 According to the atmospheric-neutrino observation
\cite{Atm}-\cite{SK1},
the lepton-flavor-mixing matrix, which we call
the Maki-Nakagawa-Sakata (MNS) \cite{MNS} matrix,
has a large mixing angle $\sin^2 2\theta_{23} \simeq 1$, 
where $\theta_{23}$ is the mixing angle between the second and the third
generations.
 This is in a clear contrast with the Cabibbo-Kobayashi-Maskawa (CKM)
\cite{CKM} quark-flavor-mixing matrix which 
does not exhibit such large mixings.
 In particular, $|V_{cb}|$ is smaller than $|V_{us}|$.
 The mixing angle between the second and the third generation is $O(1)$
in the lepton sector,
whereas it is O($10^{-2}$) in the quark sector.
 This imbalance may be a clue to obtain the
theory of flavor.
 Many attempts have been made to explain this
imbalance \cite{up2now}\nocite{Babu12}-\cite{Yanagida}.

 In the SU(5) GUT \cite{SU5,SU52}, 
the Yukawa matrix of the charged leptons ($\he$) 
is related to that of the down-type quarks ($\hd$).
 As a consequence, there are mass relations between the charged leptons
and the down-type quarks at low energies.
 The b-quark $\tau$-lepton mass ratio has been reproduced in the
original SU(5) model \cite{SU5,NOSUSY-ratio}
and in its supersymmetric version \cite{SU52,SUSY-ratio}.
 A Yukawa matrix model which reproduces all
the mass ratios between the down-type quarks and the charged leptons
has also been found within the SU(5) model \cite{GJ}. 
 However, 
the imbalance between the quark-flavor-mixing matrix 
and the lepton-flavor-mixing matrix
has not been understood within the SU(5) theory.

 In this article, we study the possibility
of naturally deriving the large flavor-mixing angle
in the lepton sector by using suitable Yukawa matrices
within the SU(5) GUT scheme.
 In particular, we examine the Fritzsch $-$ Branco $-$ Silva-Marcos (F-BS) 
type Yukawa matrices \cite{fritzsch}\nocite{BS}-\cite{no4}.
 Texture of these matrices have generic forms within the SU(5) GUT.
 It has been known \cite{no4} that the Yukawa matrices
with the F-BS texture at the GUT scale reproduce well the
observed quark masses and the CKM matrix elements.
 If we adopt the see-saw mechanism \cite{seesaw} as a natural
explanation of the tiny neutrino masses,
 we should discuss the neutrino Yukawa matrix
($\hn$) and the heavy right-handed Majorana neutrino mass matrix
(${\cal M_R}$),
in addition to the up-quark ($\hu$),
down-quark ($\hd$) and charged-lepton ($\he$) Yukawa matrices.
 The elements of the Yukawa matrices, $\hu$ and $\hd$
are constrained by the known quark masses
and the CKM matrix elements,
and those of $\he$ is related to $\hd$ in the SU(5) theory.
 Without loosing generality we can take a basis
where the Majorana mass matrix ${\cal M_R}$ is diagonal.
 The $\hn$ elements are then constrained by the observed 
neutrino mass-squared differences.
In our analysis, we assume that
the neutrinos have mass hierarchies,
$m_1 \ll m_2 \ll m_3$,
so that each $m_i$ is constrained by 
the neutrino oscillation data.
The MNS matrix elements can
then be calculated by assuming the texture of
$\hn$ at the GUT scale.
 With very simple textures of $\hn$,
the diagonal form and Fritzsch form,
we find consistent the MNS matrix elements at the weak scale,
$|U_{e2}|$ is found to be very small for the diagonal $\hn$,
whereas 
$|U_{\mu 3}| \sim 0.7$, $|U_{e 3}| < 0.2$.
 $|U_{e 2}|$ is large 
when $\hn$ has the Fritzsch form.
 Both cases are consistent with
the atmospheric neutrino-oscillation experiment and also with
the solar-neutrino experiments,
each corresponding to the small- and large-angle solution,
respectively.
 We cannot explain the LSND experiment \cite{LSND}.
 
 We are able to calculate the Jarlskog parameter of the lepton sector
($J_{\rm MNS}^{}$),
 because this parameter is related to 
that of the CKM matrix ($J_{\rm CKM}^{}$) at the GUT scale.
 In general, $J_{\rm MNS}^{}$ does not depend on the two additional
Majorana phases of the MNS matrix.
 We find that $J_{\rm MNS}^{}$ and the magnitudes of 
the MNS matrix elements are
sensitive to the $\hn$ texture at the GUT scale.

 This article is organized as follows.
 In section \ref{sec:2}, 
 we give the definition and parameterization of the MNS 
lepton-flavor-mixing matrix.
 In  section \ref{sec:2.5},
 we review recent experiments
and show allowed regions of the MNS matrix elements and
the neutrino mass differences. 
 In section \ref{sec:3}, we discuss properties of the
renormalization group equations (RGE)
for the MNS matrix elements 
and the dimension-five Majorana-mass operator.
 In section \ref{sec:3.5},
 we show the F-BS type Yukawa matrices at the GUT scale
and discuss their properties.
 In section \ref{sec:4}, 
we analyze the neutrino masses and the MNS matrix numerically
by using the 1-loop RGE of the minimal supersymmetric standard model (MSSM).
 First, we study the case where the texture of $\hn$ is diagonal,
and examine the sensitivity of the results to $M_R$,
the right-handed neutrino decoupling scale.
 Next, we study the case where $\hn$ has the 
Fritzsch form by setting $M_R$ 
at the intermediate scale, $3\times 10^{14}$ GeV.
 In section \ref{sec:sum},
 we give summary and discussions. 
\section{The Maki-Nakagawa-Sakata Matrix from the See-Saw Mechanism}
\label{sec:2}
\setcounter{equation}{0}

The Yukawa Lagrangian of leptons above the right-handed neutrino
decoupling scale ($M_R$) is{\footnote{
 Throughout this article, we use the spinor notation
where the dot product denotes the scalar product of two left-handed
Weyl spinors:
$\xi \cdot \eta = \xi^{\alpha} \epsilon_{\alpha\beta} \eta^{\beta}
= \eta \cdot \xi$
with $\epsilon_{12}=1$.
In terms of the 4-component Dirac or Majorana notation,
one finds $L \cdot e_{R}^c = \overline{e_R^{}} L$,
$L \cdot {\nu_R^c} = {\nu_R^c} \cdot L =\overline{N}L$,
${\nu_R^c}\cdot{\nu_R^c}={\overline N}{P_L}N$,
where $\overline{\psi}=\psi^{\dagger}\gamma^0$,
${\psi^c}=C{\overline{\psi}^T}$,
$P_L=(1-\gamma_5)/2$,
and $N=N^c=(\nu_R^c,\nu_R^{})^T$
is the 4-component Majorana representation of the right-handed neutrino.
}},
\bea
{\cal
L}^{\rm high}_{yukawa}
                 & = & {\he}_{ij} \phi_{\Sd} L_i \cdot e_{Rj}^c
                  +  \dfrac{1}{2}
                   \l( {\hn}_{ij} \phi_{\Su} L_i \cdot \nu_{Rj}^c
		  +    {\hn}_{ji} \phi_{\Su}\nu_{Ri}^c \cdot  L_j
	          +    {\cal M_R}_{ij}  \nu_{Ri}^c \cdot \nu_{Rj}^c \r) 
             + h.c.\,,
\nn \\
\eqlab{W_high}
\eea
where ${y^{\alpha}_{ij}}(\alpha = \nu, {\rm e})$ are
the Yukawa matrix elements of the neutrino and the charged lepton.
$\phi_{\Su}$
and $\phi_{\Sd}$
are the SU(2)$_L$ doublet Higgs bosons
that give Dirac masses to the up-type and the down-type 
fermions, respectively.
 $L_i$
is the $i$-th generation SU(2)$_L$ 
doublet lepton.
 The SU(2)$_L$ invariants are
\bea
\phi_{\Sd}L_i &=&
 \bma{c}
  \phi^0_{\Sd} \\
  \phi^{-}_{\Sd}
 \ema
 \bma{c}
  \nu_{Li} \\
  l_{Li}
 \ema
 = 
 \phi^0_{\Sd}l_{Li}-\phi^{-}_{\Sd}\nu_{Li}\,, \nn \\
\phi_{\Su}L_i &=&
 \bma{c}
  \phi^{+}_{\Su} \\
  \phi^{0}_{\Su}
 \ema
 \bma{c}
  \nu_{Li} \\
  l_{Li}
 \ema
 = 
 \phi^+_{\Su}l_{Li}-\phi^{0}_{\Su}\nu_{Li}\,.
\eea
 $\nu_{Ri}^{}$ and $e_{Ri}^{}$
are the $i$-th generation right-handed neutrino and
charged lepton, respectively.
 Majorana mass matrix ${\cal M_R}_{ij}$ is complex and symmetric.
 When the right-handed neutrinos $\nu_{Ri}^{}$ are
integrated out at the scale $M_R$,
the effective Lagrangian becomes
\beq
{\cal
L}^{\rm low}_{yukawa}
           =  {\he}_{ij} \phi_{\Sd} L_i \cdot e_{Rj}^c
            - \dfrac{1}{2} \kappa_{ij}(\phi_{\Su}L_i)\cdot(\phi_{\Su}L_j)
            + h.c.\,,
\eqlab{W_low}
\eeq
with
\beq
\kappa =   \hn {\cal M_R}^{-1} \hn^{T}\,.
\eqlab{kappa-0}
\eeq
The charged-lepton and neutrino
mass-matrices are obtained as
\beq
M_e^{\ast}   = - \he \VEV{\phi_{\Sd}^0}\,, \mbox{\quad\quad}
M_\nu^{\ast} = \kappa \l(\VEV{\phi_{\Su}^0}\r)^2\,.
\eeq
The lepton mass terms can be expressed as
\bea
{\cal L}_{mass}
&=& -\l(M_e\r)^{\ast}_{ij} l_{Li}\cdot e_{Rj}^c
    - \dfrac{1}{2} \l(M_\nu\r)^{\ast}_{ij} \nu_{Li}^{}\cdot \nu_{Lj}^{}
    +h.c. \nn \\
&=& -\l(M_e\r)^{\ast}_{ij} \overline{l_{Rj}}l_{Li}
    - \dfrac{1}{2} \l(M_\nu\r)^{\ast}_{ij} \overline{n_{Ri}^{}} n_{Lj}^{}
    +h.c. \nn \\
&=& - \overline{l_L} M_e l_R - \overline{l_R} M_e^{\dagger} l_L
    - \dfrac{1}{2} \l(\overline{n_L^{}}M_\nu n_R^{}
                     +\overline{n_R^{}}M_\nu^{\dagger} n_L^{} \r)\,,
\eea
where we introduce the 4-component Majorana field for the light neutrinos,
$n_i^{} \equiv \l(\nu_{Li}^{},\nu_{Li}^c\r)^T$.
$M_e$ is a general complex matrix, whereas $M_\nu$ is a symmetric
complex matrix in the generation space.

 We give the definition and useful parameterization of the $3 \times 3$
Maki-Nakagawa-Sakata (MNS) lepton-flavor-mixing matrix \cite{MNS}.
 The MNS matrix is defined analogously to the 
CKM matrix \cite{CKM,PDG}, 
in terms of the unitary matrices $U_{\Se}$ and $U_{\nu}$
that transform the mass-eigenstates into the
weak-current eigenstates:
\beq
\bma{c}
l_{L1}^{} \\
l_{L2}^{} \\
l_{L2}^{}
\ema
 =
U_{\Se}
\bma{c}
e_{L}^{} \\
\mu_{L}^{} \\
\tau_{L}^{}
\ema
\,,~~~~~~~~~~
\bma{c}
\nu_{L1}^{} \\
\nu_{L2}^{} \\
\nu_{L3}^{}
\ema
 =
U_{\nu}
\bma{c}
\nu_1^{} \\
\nu_2^{} \\
\nu_3^{}
\ema
\,.
\eqlab{UeUn}
\eeq
 That is, these unitary matrices diagonalize 
the  charged-lepton mass-matrix squared 
\beq
U_{\Se}^{\dagger} M_{\Se} M_{\Se}^{\dagger}U_{\Se}
=  diag.(m_e^2,m_\mu^2,m_\tau^2)\,,
\eeq
and the Majorana-neutrino mass-matrix
\beq
U_{\nu}^{T} M_{\nu}^{\dagger} U_{\nu} = diag.(m_1^{}, m_2^{}, m_3^{})\,,
\eeq
where $0 \leq m_1 \ll m_2 \ll m_3$.
The MNS matrix is then defined as:
\beq
\eqlab{def_MNS}
\l(V_{\rm MNS}^{}\r)_{\alpha i} \equiv
\l( U_{\Se}^{\dagger} U_{\nu}^{} \r)_{\alpha i}^{}\,,
\mbox{\quad \quad}
\nu_\alpha^{}=\sum^3_{i=1} \l(V_{\rm MNS}^{}\r)_{\alpha i}^{} \nu_i^{}\,,
\eeq
where $\alpha$ and $i$ label the neutrino flavors
($\alpha=e, \mu, \tau$)
and the mass eigenstates ($i=1,2,3$).
 In terms of the flavor-state $\nu_\alpha^{}$ ($\alpha=e, \mu, \tau$),
the leptonic charged-current interactions take
the flavor-diagonal form
\bea
{\cal L}_{CC} 
&=& -\dfrac{g}{\sqrt{2}} W^{-}_{\mu}
\sum_{l=e,\mu,\tau} \overline{l_{L}^{}} \gamma^{\mu} \nu_{l}^{}
+h.c.\,.
\eea 

The $3\times3$ MNS matrix has three mixing angles
and three phases in general.
We adopt the following parameterization
\beq
V_{\rm MNS}^{} =
\bmaT
U_{e 1}    & U_{e 2}    & U_{e 3} \\
U_{\mu 1}  & U_{\mu 2}  & U_{\mu 3} \\
U_{\tau 1} & U_{\tau 2} & U_{\tau 3}  
\emaT
\bmaT
1 & 0 & 0 \\
0 & e^{i \varphi_2^{}} & 0 \\
0 & 0 & e^{i \varphi_3^{}} 
\emaT\,,
\eqlab{MNS1}
\eeq
 where the two Majorana phases, $\varphi_2^{}$ and $\varphi_3^{}$,
are given explicitly.
 The remaining matrix $U$,
which has three mixing angles and one phase,
can be parameterized the same way as the CKM matrix.
 Because the present neutrino oscillation experiments constrain
directly the elements, $U_{e2}$, $U_{e3}$, and $U_{\mu 3}$,
we find it most convenient to adopt the parameterization \cite{Wolf,MK}
where we take these three matrix elements
in the upper-right corner of the matrix
as the independent parameters.
 Without losing generality, we can take $U_{e2}$
and $U_{\mu 3}$ to be real and non-negative.
 By allowing $U_{e3}$ to have a complex phase,
$U_{e3}=|U_{e3}| e^{i \delta_{\rm MNS}^{}}$,
the number of the independent parameters is four.
 All the other matrix elements are them determined by the
unitary conditions:
\bseq
\bea
U_{e1} &=& \sqrt{1-|U_{e3}|^2-|U_{e2}|^2}\,, \eqlab{MNS-R1}\\ 
U_{\tau
3}
&=&
\sqrt{1-|U_{e3}|^2-|U_{\mu
3}|^2}\,,\eqlab{MNS-R2} 
\\ 
U_{\mu 1} &=& - \dfrac{U_{e2}U_{\tau 3} + U_{\mu 3}U_{e1}U_{e3}^{\ast} }
                      {1-|U_{e3}|^2}\,, \eqlab{MNS-C1}\\
U_{\mu 2} &=& \dfrac{U_{e1}U_{\tau 3} - U_{\mu 3}U_{e2}U_{e3}^{\ast} }
                    {1-|U_{e3}|^2}\,, \eqlab{MNS-C2}\\
U_{\tau 1} &=& \dfrac{U_{e2}U_{\mu 3} - U_{\tau 3}U_{e1}U_{e3}^{\ast} }
                     {1-|U_{e3}|^2}\,, \eqlab{MNS-C3}\\
U_{\tau 2} &=& - \dfrac{U_{\mu 3}U_{e1} + U_{e2}U_{\tau 3}U_{e3}^{\ast} }
                       {1-|U_{e3}|^2}\,.\eqlab{MNS-C4}
\eea
\eqlab{MNS-RC}
\eseq
\hspace*{-1ex}
In this parameterization,
$U_{e1}$, $U_{e2}$, $U_{\mu 3}$, and $U_{\tau 3}$
are real and non-negative numbers,
and the other elements are complex numbers.
In particular, we note
\beq
\l|\l(V_{\rm MNS}^{} \r)_{e2}^{}\r| = U_{e2}\,, 
\mbox{\quad \quad}
\l|\l(V_{\rm MNS}^{} \r)_{\mu 3}^{}\r| = U_{\mu 3}.
\eeq
The two Majorana phases $\varphi_2^{}$ and $\varphi_3^{}$
do not contribute to the Jarlskog parameter \cite{JarP}
of the MNS matrix:
\bea
J_{\rm MNS}^{}
&=& Im \l({V_{e1} V_{\tau 1}^{\ast} V_{\tau 3} V_{e3}^{\ast}}\r) \nn \\
&=& Im \l({U_{e1} U_{\tau 1}^{\ast} U_{\tau 3} U_{e3}^{\ast}}\r) \nn \\
&=& \dfrac{U_{e1}U_{e2}U_{\mu 3}U_{\tau 3}}{1-\l|U_{e3}\r|^2}
              Im \l({U_{e3}}\r)\,,
\eqlab{Jya-Lepton}
\eea
where $Im \l({U_{e3}}\r) = |U_{e3}| \sin\l(\delta_{\rm MNS}^{}\r)$.
\section{Experimental constraints}
\label{sec:2.5}
\setcounter{equation}{0}
 In this section we review briefly the experimental constraints 
on the neutrino masses and the MNS matrix elements
under the assumptions of the three neutrino flavors
and the mass hierarchy $m_1^{} \ll m_2^{} \ll m_3^{}$.
\subsection{Constraints on the MNS matrix elements}
 The survival and transition probabilities of the neutrino oscillation
in the vacuum
take the following simple form \cite{prob},
\bea
\eqlab{prob}
P_{\nu_\alpha^{} \to \nu_\beta^{}} &=& 
4 \l| U_{\alpha 3} \r|^2 \l| U_{\beta 3} \r|^2
\sin^2 \l({\dfrac{m_3^2-m_2^2}{4E}}L\r)\,, \\
\eqlab{surv}
P_{\nu_\alpha \to \nu_\alpha} &=& 1 - 
4 \l| U_{\alpha 3} \r|^2
\l(1- \l| U_{\alpha 3} \r|^2 \r)
\sin^2 \l({\dfrac{m_3^2-m_2^2}{4E}}L\r)\,,
\eea
when the following condition is satisfied{\footnote
{See Appendix A for more details.}}:
\bea
\eqlab{massd2}
\dfrac{m_2^2-m_1^2}{2E}L \ll 1 \sim \dfrac{m_3^2-m_2^2}{2E}L\,.
\eea
 On the other hand,
the probabilities take the following form,
\bea
P_{\nu_{\alpha} \to \nu_{\beta}}
&=&
2|U_{\alpha 3}|^2|U_{\beta 3}|^2 
\nn \\& &
-
\l[
4 Re(U_{\alpha 1}U_{\beta 1}^{\ast}U_{\beta 2}U_{\alpha 2}^{\ast})
\sin^2\l(\dfrac{m_2^2-m_1^2}{4E}L\r)
+2 J_{\rm MNS}
\sin^2\l(\dfrac{m_2^2-m_1^2}{2E}L\r)\r]\,,
\nn \\
\eqlab{prov2} \\
P_{\nu_\alpha \to \nu_\alpha} &=&
1
-2 |U_{\alpha 3}|^2 \l(1-|U_{\alpha 3}|^2\r)
-4 |U_{\alpha 1}|^2 |U_{\alpha 2}|^2 
\sin^2
\l(
{\dfrac{\delta m_{12}^2}{4E}L}
\r)\,.
\eqlab{A9-2}
\eea
when the following conditions are satisfied:
\bea
\eqlab{massd3}
\dfrac{m_2^2-m_1^2}{2E}L \sim 1 \ll \dfrac{m_3^2-m_2^2}{2E}L\,.
\eea
Below, we obtain constraints on the neutrino mass differences
and the MNS matrix elements from the recent neutrino-oscillation
experiments with the above approximations.

\subsubsection{CHOOZ experiments}
 The CHOOZ experiment \cite{CHOOZ}
measured the survival probability of $\overline{\nu_e^{}}$,
and found 
\beq
\sin^2 2 \theta_{\rm CHOOZ}
< 0.18\,,
~~~\mbox{for}~~~ \delta m_{\rm CHOOZ}^2 > 1 \times 10^{-3} \mbox{eV}^2.
\eqlab{chooz0}
\eeq
By assuming $m_2^2-m_1^2 \ll 1 \times 10^{-3}$ eV$^2$ to accommodate
the solar-neutrino oscillation (see below),
we can use \eqref{surv} to obtain the following constraint:
\bseq
\bea
\l | U_{e3} \r|^2 \l( 1-\l| U_{e3} \r|^2 \r)
&<& 0.045\,,
\eqlab{region} \\
\mbox{for  \quad}
m_3^2-m_1^2 \simeq m_3^2-m_2^2 
&>& 1 \times 10^{-3} \mbox{eV}^2\,.
\eqlab{CHOOZ-MASS}
\eea
\eseq

\subsubsection{Solar-neutrino deficit}
Deficits of solar neutrinos observed at several telestial experiments
\cite{solar3,solar4}
have been successfully interpreted in terms of the
$\nu_e^{}\to\nu_X^{}$ ($\nu_X^{}\neq\nu_e^{}\,,\overline{\nu_e^{}}$)
oscillation in the following three scenarios \cite{solar}.

 {\sf MSW small-mixing solution}:
\bseq
\bea
\eqlab{MSW-S}
3 \times 10^{-3} < &\sin^2 2\theta_{\rm SUN}& < 1.1 \times 10^{-2}\,, \\
\mbox{and \quad}
4\times10^{-6} < &\delta m_{\rm SUN}^2 
(\mbox{eV}^2)& < 1.2 \times 10^{-5}\,,
\eea
\eseq

 {\sf MSW large-mixing solution}:
\bseq
\bea 
\eqlab{MSW-L}
0.42 < &\sin^2 2\theta_{\rm SUN}& < 0.74\,, \\
\mbox{and \quad}
8\times10^{-6} < &
\delta m_{\rm SUN}^2 (\mbox{eV}^2)
 & < 3.0 \times 10^{-5}\,,
\eea
\eseq

 {\sf Vacuum oscillation solution}:
\bseq
\bea 
\eqlab{Vacuum}
0.7 < &\sin^2 2\theta_{\rm SUN}& < 1.0\,, \\
\mbox{and \quad}
6\times10^{-11} < &
\delta m_{\rm SUN}^2 
(\mbox{eV}^2)& < 1.1 \times 10^{-10}\,.
\eea
\eseq

Under the mass-hierarchy assumption of
$m_1 \ll m_2 \ll m_3$,
the solar-neutrino deficits should be explained by the oscillation of
the lighter two neutrinos
in both the MSW \cite{MSW} and the vacuum-oscillation \cite{vacuum}
scenarios.
The $\nu_e^{}$ survival probability in the vacuum can then 
be expressed by \eqref{A9-2}.
Because the mass-squared difference $m_3^2-m_1^2$ suggested
by the atmospheric neutrino observation is
much larger than the differences \eqref{MSW-S}$\sim$\eqref{Vacuum},
there appears an energy-independent deficit factor
$-2|U_{e3}|^2\l(1-|U_{e3}|^2\r)$.
 This factor should be smaller than $9\%$ by the CHOOZ experiment
if $m_3^2-m_1^2 > 1 \times 10^{-3}$ eV$^2$; see \eqref{CHOOZ-MASS}.
 It is also constrained by the observation of lower energy
solar neutrino \cite{solar4}.
 Since we need only rough estimate of the allowed ranges
of the MNS matrix elements\footnote{
{Quantitative study of three-flavor oscillation effects will be 
reported elsewhere.}},
we ignore the small energy-independent deficit factor
and interpret the results of the two-flavor
analysis \eqref{MSW-S} $\sim$ \eqref{Vacuum}
by using the following identifications:
\bseq
\bea
\sin^2 2\theta_{\rm SUN} &=& 4 \l|U_{e1}\r|^2\l|U_{e2}\r|^2 \nn \\
&=&4\l( 1 -\l|U_{e2}\r|^2-\l|U_{e3}\r|^2 \r)\l|U_{e2}\r|^2 \,,\\
\delta m^2_{\rm SUN} &=& m_2^2-m_1^2\,.
\eea
\eseq

\subsubsection{Atmospheric-neutrino anomaly}
The recent analysis of the atmospheric neutrino data
from the Super-Kamiokande experiment finds \cite{Atm}.
\bseq
\bea
0.7 < &\sin^2 2\theta_{\rm ATM}& < 1\,,
\eqlab{atm2} \\
\mbox{and \quad}
3\times10^{-4} < &\delta m^2_{\rm ATM}
(\mbox{eV}^2)
& < 7 \times 10^{-3}\,,
\eqlab{atm0}
\eea
\eseq
for $\nu_\mu^{}\to\nu_X^{}$ ($\nu_X^{} \neq \nu_\mu^{}, \nu_e^{}$)
oscillation.
The $\nu_\mu^{}\to\nu_e^{}$ oscillation scenario is not
only disfavored by the CHOOZ experiment \eqref{chooz0},
but also disfavored by the Super-Kamiokande datum by itself.
In our three-flavor analysis, the data should be interpreted
with the $\nu_\mu \to \nu_\tau$ oscillation under the constraint
\eqref{massd2}.
By using \eqref{surv}, we have the following identifications:
\bseq
\bea
\sin^2 2\theta_{\rm ATM} &=& 
4 |U_{\mu 3}|^2 \l(1-|U_{\mu 3}|^2\r)\,, \\
\delta m^2_{\rm ATM} &=& m_3^2-m^2_2\,.
\eea
\eseq

\subsubsection{Neutrino masses}
 The neutrino oscillation experiments measure
only the mass-squared differences of the three neutrinos.
Under the assumption of the neutrino-mass hierarchies,
\beq
m_1 \ll m_2 \ll m_3\,,
\eqlab{massd1}
\eeq
the absolute values of the neutrino masses can be constrained
determined as follows:
\bseq
\bea
m_3^2 &\simeq& m_3^2-m_2^2 = \delta m^2_{\rm ATM}\,, \\
m_2^2 &\simeq& m_2^2-m_1^2 = \delta m^2_{\rm SUN}\,.
\eea
\eseq
The heaviest neutrino mass is then determined by \eqref{atm0}
\beq
m_3 = (0.02 \sim 0.08) \mbox{ eV}\,, 
\eqlab{mass_3}
\eeq
and there are three possibilities for $m_2$:
\bseq
\bea
m_2 &=& (0.002 \sim 0.003) {\mbox { eV}} \nn \\
& &\mbox{~~~~~~({\sf MSW small-mixing solution})}\,, \eqlab{mass_2_MSWS} \\
m_2 &=& (0.003 \sim 0.005) {\mbox { eV}} \nn \\
& &\mbox{~~~~~~({\sf MSW large-mixing solution})}\,, \eqlab{mass_2_MSWL} \\
m_2 &=& (8\sim10) \times 10^{-6} {\mbox { eV}} \nn \\
& &\mbox{~~~~~~({\sf Vacuum mixing solution})}\,.\eqlab{mass_2_Vac}
\eea
\eseq
$m_1$ cannot be determined.
Since its magnitude plays little role in the following
analysis, we simply assume the hierarchy \eqref{massd1}:
\bea
m_1^2 < \dfrac{m_2^2}{10}\,.
\eqlab{mass_1}
\eea

\subsection{Quark and lepton masses and the CKM matrix}
In our analysis, we adopt the following 
quark and lepton masses at the Z-boson mass scale \cite{PDG,masses}
to constrain the Yukawa matrices elements:
\bea
m_t^{} (m_Z^{}) & = & 175    \pm 6     {\mbox {\hspace{11.0ex} GeV}}\,, \nn \\
m_c^{} (m_Z^{}) & = & 0.59   \pm 0.07   {\mbox {\hspace{7.5ex} GeV}}\,, \nn \\
m_u^{} (m_Z^{}) & = & 0.0022 \pm 0.0007 {\mbox {\hspace{3ex} GeV}}\,,
\eqlab{up-part}
\eea
\bea
m_b^{} (m_Z^{}) & = & 3.02   \pm 0.19   {\mbox {\hspace{7.5ex} GeV}}\,, \nn \\
m_s^{} (m_Z^{}) & = & 0.077  \pm 0.011  {\mbox {\hspace{5.3ex} GeV}}\,, \nn \\
m_d^{} (m_Z^{}) & = & 0.0038 \pm 0.0007 {\mbox {\hspace{3ex} GeV}}\,,
\eqlab{down-part} 
\eea
\bea
m_\tau^{} (m_Z^{}) & = & 1746.5 \pm 0.3 {\mbox {\hspace{7.1ex} MeV}}\,, \nn \\
m_\mu^{} (m_Z^{})  & = & 102.7          {\mbox {\hspace{14ex} MeV}}\,, \nn \\
m_e^{} (m_Z^{})    & = & 0.487          {\mbox {\hspace{14ex} MeV}}\,.
\eqlab{charged-lepton} 
\eea

The CKM matrix elements 
are determined by the magnitudes of the
three off-diagonal elements \cite{PDG}
\bseq
\bea
\l|V_{us}\r| &=& 0.219 \sim 0.224\,, \\
\l|V_{cb}\r| &=& 0.036 \sim 0.046\,, \\
\dfrac{\l|V_{ub}\r|}{\l|V_{cb}\r|} &=& 0.06 \sim 0.10\,,
\eea
\eqlab{exp_ckm}
\eseq
\hspace*{-1ex}
and one phase,
which in our parameterization is related to the Jarlskog parameter as
\beq
\sin \delta_{\rm CKM}^{}
 = \dfrac{\l(1-|V_{ub}|^2\r)}{V_{ud}V_{us}V_{cb}V_{tb}|V_{ub}|}
J_{\rm CKM}^{}\,,
\eeq
where $\delta_{\rm CKM}^{}$ is the $CP$ violation angle.
We adopt the same phase combination for the MNS $(U_{\alpha i})$
and the CKM matrices, and
$V_{ud}$, $V_{us}$, $V_{cb}$, $V_{tb}$ are all real and positive.
The remaining CKM matrix elements are determined by the
unitarity conditions analogously to \eqref{MNS-RC} \cite{MK}.

Constraints on the neutrino masses and the MNS matrix elements
depend on the three scenarios that explain the solar-neutrino
experiments.
 In \Tbref{MNS-Mass} we summarize the typical
allowed ranges of these parameters
under the assumption of $m_1 \ll m_2 \ll m_3$
for the three scenarios. We denote 
``MSW-S'' for the MSW small-mixing scenario,
``MSW-L'' for the MSW large-mixing scenario, and
``V-O'' fot the vacuum-oscillation scenario.
\begin{table}[htbp]
\begin{center}
\begin{tabular}[tb]{|l||c|c||c|c|c|}
\hline
      & $m_2$ (eV) & $m_3$ (eV)
      & $|U_{e2}|$ & $|U_{e3}|$ & $|U_{\mu 3}|$ \\
\hline
\hline
MSW-S & $0.002 \sim 0.003$ & $0.02 \sim 0.08$
      & $0.027 \sim 0.053$ & $<0.22$ & $0.48 \sim 0.88$ \\
\hline
MSW-L & $0.003 \sim 0.005$ & $0.02 \sim 0.08$
      & $0.35 \sim 0.50$ & $<0.22$ & $0.48 \sim 0.88$ \\
\hline
V-O   & $(8 \sim 10) \times 10^{-6}$ & $0.02 \sim 0.08$
      & $0.48 \sim 0.71$ & $<0.22$ & $0.48 \sim 0.88$ \\
\hline
\end{tabular}
\caption{Allowed ranges of the neutrino masses and the MNS matrix elements 
in the three scenarios under the assumption $m_1 \ll m_2 \ll m_3$.}
\tblab{MNS-Mass}
\end{center}
\end{table}
\section{Renormalization Group Equations}
\label{sec:3}
\setcounter{equation}{0}

 The RGE of the coefficient $\kappa$ of dimension-five operator 
in the effective Lagrangian \eqref{W_low},
which is formed by the see-saw mechanism \cite{seesaw},
has been studied in 
Refs. \cite{nue-RGE}.
Below the right-handed neutrino mass scale $M_R$,
the matrix $\kappa$ satisfies the following RGE in the MSSM \cite{nue-RGE}:
\beq
8\pi^2\dfrac{d}{dt}\kappa =
      \l\{ \tr \l( 3\hu\hu^{\dagger} \r) 
       - 4\pi \l( 3\alpha_2 + \dfrac{3}{5} \alpha_1 \r) \r\} \kappa
+\dfrac{1}{2} \l\{ \l( \he \he^{\dagger}\r) \kappa
                 + \kappa\l( \he \he^{\dagger}\r)^{T} \r\}\,,
\eqlab{RGE_kappa}
\eeq
where $t=\ln \mu$, $\hu$ is the up-quark Yukawa matrix and $\he$ is the
charged-lepton Yukawa matrix.
 The RGE's of these matrices and the gauge couplings are summarized in
Appendix B.
 When the contributions from the first-generation
fermions can be neglected,
the RGE of the lepton-flavor mixing angle $\theta_{23}$
($|U_{\mu 3}|\sim \sin \theta_{23}$)
takes a very simple form \cite{nue-RGE}:
\beq
8\pi^2 \dfrac{d}{dt} \sin^2{2\theta_{23}}
=
- \sin^2{2\theta_{23}} \l( 1-\sin^2{2\theta_{23}} \r)
\l( |y_\tau|^2-|y_\mu|^2 \r)
\dfrac{\kappa_{33}+\kappa_{22}}{\kappa_{33}-\kappa_{22}}\,,
\eqlab{RGE_theta}
\eeq
where $y_\mu$ and $y_\tau$ are the second and third generation
charged-lepton Yukawa couplings, respectively.
 It should be noted in \eqref{RGE_theta} that
if $\kappa_{33}>\kappa_{22}$ holds,
then $\sin^2 2\theta_{23}=1$ is the infra-red fixed point of the
equation \cite{nue-RGE}, and hence it is possible to obtain 
$\sin^2 2\theta_{23} \sim 1$ at the $m_Z^{}$ scale
as long as its magnitude is large at the 
right-handed neutrino decoupling scale, $M_R$.

\section{GUT scale Yukawa matrix texture}
\label{sec:3.5}
\setcounter{equation}{0}
 The Yukawa terms at the SU(5) GUT scale are
\bea
{\cal L}_{yukawa}^{\rm GUT}
 &=& y^{10}_{ij} \varepsilon^{abcde}
                 (\phi^{\Su})_a (\chi_i)_{bc} \cdot (\chi_j)_{de}
            + y^{5^{\ast}}_{ij} (\phi^{\Sd})^a (\chi_i)_{ab} \cdot (\psi_j)^b
\nn \\
	   & &+ y^{1}_{ij} ({\phi^{\Su}})_a {(\psi_i)^a}\cdot ({\nu_R^c}_j^{})
              + {\cal M_R}_{ij} ({\nu_R^c}_i^{}) \cdot ({\nu_R^c}_j^{})
                + h.c.\,,
\eqlab{SU5-yukawa}
\eea
where $\chi$, $\psi$ and $\nu_R^c$
are all left-handed fermions which transform as
${\bf 10}$, ${\bf 5^{\ast}}$ and
${\bf 1}$ representation of the SU(5) group, respectively.
 Indices $i,j$ ($=1\sim3$) stand for generations and 
$a,b,\dots$ ($=1\sim5$) give the SU(5) indices.
$\phi^{\Su}$ and $\phi^{\Sd}$ 
are now ${\bf 5}$ and ${\bf 5^{\ast}}$ representation Higgs
particles, respectively.
The Yukawa matrices of the SM are related to those at the GUT scale as,
\bseq
\bea
{\hu}_{ij} = {\hu}_{ji} &=& y^{10}_{ij}\,, \eqlab{yuGUT}\\
{\hd}_{ij} = {\he}_{ji} &=& y^{5^{\ast}}_{ij}\,,\eqlab{ydGUT} \\
{\hn}_{ij} &=& y^{1}_{ij} \eqlab{ynGUT}\,.
\eea
\eseq
We note here that the Yukawa matrix $\hu$ is symmetric and that
the Yukawa matrix $\he$ is the transpose on the
Yukawa matrix $\hd$ in the flavor space:
\beq
\hu = \l(\hu\r)^T\,,
~~~~~~~~~~~
\he = (\hd)^T\,,
\eeq 
at the GUT scale in the SU(5) theory \cite{up2now2}.
These are direct consequences of the SU(5)
representations of quarks and leptons:
\bea
{\bf 10}       &:& {\chi_{ab}} = {\rm u}^c + {\rm Q} + {\rm e}^c\,, \nn \\
{\bf 5^{\ast}} &:& {\psi^a} = {\rm d}^c + {\rm L}\,, \nn \\
{\bf 1}        &:& \nu^c_R\,,
\eqlab{repSU5}
\eea
where ${\rm u}^c$ and ${\rm d}^c$
are the charge-conjugation of the right-handed
up and down quarks, respectively.
${\rm Q}$ is the SU(2)$_L$ doublet left-handed quarks.

 Both up- and down-type Yukawa matrices can be transformed
into the nearest-neighbor-interaction (NNI) 
form by a weak-basis transformation 
without loosing generality \cite{NNI}. 
 Recently, many authors studied the texture of the Yukawa matrices
at the $m_Z^{}$ scale in the NNI basis \cite{ito}\nocite{no1}-\cite{no2}.
 E.~Takasugi has further shown \cite{Takasugi} that 
one of the up- or down-quark mass matrices can be transformed
into either
the Fritzsch form \cite{fritzsch} or the BS form \cite{BS},
while keeping the other matrix in the NNI form.
For instance, under the SU(5) constraint, \eqref{yuGUT},
we can take the Fritzsch texture for the up-quark Yukawa matrix,
\beq
{\overline {\rm Q}_L} \hu {\rm u}_R = {\overline {\rm Q}_L}
\bmaT
   0    & a_{\Su} & 0      \\
a_{\Su} &    0    & b_{\Su}\\
   0    & b_{\Su} & c_{\Su}
\emaT
{\rm u}_R\,,
\eqlab{up-1}
\eeq
 where $a_{\Su}$ and $b_{\Su}$ are real numbers
and $c_{\Su}$ is a complex number
without loosing generality.
 The down-quark Yukawa matrix $(\hd)$ can still
be parameterized by the NNI form,
\beq
{\overline {\rm Q}_L} \hd {\rm d}_R = {\overline {\rm Q}_L}
\bmaT
   0   & x_{12} & 0      \\
x_{21} &    0   & x_{23}\\
   0   & x_{32} & x_{33}
\emaT
{\rm d}_R\,,
\eqlab{down-0}
\eeq
 in general.
 Here the nonzero five $x_{ij}$ components are complex numbers.
 Three phases in \eqref{down-0} can be removed by
using the rephasing freedom of ${\rm d}_R$,
and $c_{\Su}$ can be made real by rephasing ${\rm u}_{R}$ once the
GUT is broken.
The parameterization is still most general,
containing 10 real parameters for six quark masses and 
four independent elements of the CKM matrix.

 If we adopt the parameterization of \eqref{up-1} at the weak scale,
three parameters are given analytically in terms of 
the up-quark-mass eigenvalues\footnote{{
When $a_{\Su}$, $b_{\Su}$, and $c_{\Su}$ are non-negative real numbers,
the determinant of $\hu$ is negative.
On the other hand, all up-quark-mass eigenvalues are real and positive.
These two statements are consistent,
because we can choose the unitary matrix for the right-handed fields as
$\det (\hu) \det\l({U_R^{\Su}}\r) v_{\Su}^3 = m_u m_c m_t$,
with $\det \l({U_R^{\Su}}\r)=-1$.
}}:
\beq
m_u^{} = \dfrac{v_{\Su}}{\sqrt{2}} \dfrac{a_{\Su}^2 c_{\Su}}{b_{\Su}^2}\,,
\mbox{\hspace{5ex}}
m_c^{} = \dfrac{v_{\Su}}{\sqrt{2}} \dfrac{b_{\Su}^2}{c_{\Su}}\,,
\mbox{\hspace{5ex}}
m_t^{} = \dfrac{v_{\Su}}{\sqrt{2}}  c_{\Su}\,.
\eqlab{mu-1}
\eeq
Here $v_{\Su}/\sqrt{2}=v/\sqrt{2} \cdot \sin \beta(>0)$
is the vacuum expectation value (vev) of the $\phi_{\Su}^0$ field,
where 
\beq
\tan \beta = \dfrac{\VEV{\phi_{\Su}^0}}{\VEV{\phi_{\Sd}^0}}
=\dfrac{v_{\Su}}{v_{\Sd}}\,,
\eeq
and $v_{\Su}^2+v_{\Sd}^2=v^2 \simeq$ (246 GeV)$^2$.
The observed up-quark masses at the $m_Z^{}$ scale implies the hierarchy
\beq
a_{\Su}:b_{\Su}:c_{\Su} =
\dfrac{\sqrt{m_u m_c}}{m_t}:\sqrt{\dfrac{m_c}{m_t}}:1\,.
\eeq
On the other hand, the well-known empirical relations
at the $m_Z^{}$ scale,
\beq
\l| V_{us} \r| \sim \sqrt{\dfrac{m_d}{m_s}}\,,  
\mbox{\hspace {5ex}}
\l| V_{cb} \r| \sim \dfrac{m_s}{m_b}\,,
\mbox{\hspace {5ex}}
\l| \dfrac{V_{ub}}{V_{cb}} \r| \sim \sqrt{\dfrac{m_u}{m_c}}\,,
\eeq
 implies in the NNI basis of \eqref{down-0} that the $\hd$ elements
should satisfy \cite{Takasugi2}
\beq
|x_{12}| \sim |x_{21}| \mbox{\quad and \quad} |x_{32}| \sim |x_{33}|\,.
\eeq
 Hence at the weak scale, the matrix $\hd$
should approximately have the Branco$-$Silva-Marcos (BS) 
form \cite{BS},
\beq
{\overline {\rm Q}_L} \hd {\rm d}_R = {\overline {\rm Q}_L}
\bmaT
   0    & a_{\Sd}e^{i\phi_1} & 0      \\
a_{\Sd}e^{-i\phi_1} &    0    & b_{\Sd}e^{i\phi_2}\\
   0    & c_{\Sd} & c_{\Sd}
\emaT
{\rm d}_R\,,
\eqlab{down-1}
\eeq
 where $a_{\Sd}$, $b_{\Sd}$ and $c_{\Sd}$ are real positive numbers and 
$\phi_{1,2}$ are phases.
 We have made use of the rephasing freedom to obtain the
above phase assignment. 
 These textures of $\hu$ and $\hd$ are one of the simplest set of 
the quark Yukawa matrices at the $m_Z^{}$ scale,
which is consistent with all the experimental data.
The down-quark mass eigenvalues are then related to the parameters
$a_{\Sd}$, $b_{\Sd}$ and $c_{\Sd}$ of
\eqref{down-1} as
\beq
m_d^{} = \dfrac{v_{\Sd}}{\sqrt{2}} \dfrac{a_{\Sd}^2}{b_{\Sd}}\,,
\mbox{\hspace{5ex}}
m_s^{} = {v_{\Sd}}\dfrac{b_{\Sd}}{2}\,,
\mbox{\hspace{5ex}}
m_b^{} = {v_{\Sd}} c_{\Sd}\,,
\eqlab{md-1}
\eeq
where $v_{\Sd}/\sqrt{2}=v/\sqrt{2} \cdot \cos \beta (>0)$
is the vev of the $\phi_{\Sd}^0$ field.

 Because the renormalization effect is not overwhelming
between the GUT scale and weak scale in the MSSM \cite{no4},
we adopt the parameterization \eqref{up-1} for $\hu$
and \eqref{down-1} for $\hd$ at the GUT scale.
 The elements \eqref{mu-1} and \eqref{md-1} are then
valid for effective ``masses'' at the GUT scale.
 The charged-lepton Yukawa matrix $\he$ is the
transpose of the down-quark one $\hd$ at the GUT scale.
The asymmetry of the BS texture then leads to the sharp difference
between the down-quark and the charged-lepton sectors in the
squared Yukawa matrices
that are diagonalized by the relevant Unitary matrices:
\bseq
\bea
\hd\hd^{\dagger}
&=&
\bmaT
|a_{\Sd}|^2 &            0            & a_{\Sd} c_{\Sd} e^{i \phi_1} \\
0            & |a_{\Sd}|^2+|b_{\Sd}|^2 & b_{\Sd} c_{\Sd} e^{i \phi_2} \\
a_{\Sd} c_{\Sd} e^{-i \phi_1} & b_{\Sd} c_{\Sd} e^{-i \phi_2} &2c_{\Sd}^2
\emaT
\simeq
\dfrac{m_b^2}{v_{\Sd}^2}
\bmaT
|\alpha_{\Sd}|^2 &          0     & \alpha_{\Sd} \\
0              & |\beta_{\Sd}|^2  & \beta_{\Sd} \\
\alpha_{\Sd}^{\ast}   & \beta_{\Sd}^{\ast}    & 2
\emaT\,,
\eqlab{ydyd}
\\
\nn \\
\he\he^{\dagger}
&=&
\bmaT
|a_{\Sd}|^2  &    0    & a_{\Sd}b_{\Sd} e^{-i({\phi_1+\phi_2})} \\
0            & a_{\Sd}^2 + c_{\Sd}^2   & c_{\Sd}^2\\
 a_{\Sd}b_{\Sd} e^{i({\phi_1+\phi_2})} & c_{\Sd}^2 & b_{\Sd}^2+c_{\Sd}^2  
\emaT
\simeq
\dfrac{m_\tau^2}{v_{\Sd}^2}
\bmaT
|\alpha_{\Se}|^2                      & 0  & \alpha_{\Se}\beta_{\Se} \\
0                                     & 1+|\alpha_{\Se}|^2  & 1 \\
\alpha_{\Se}^{\ast}\beta_{\Se}^{\ast} & 1  & 1+|\beta_{\Se}|^2
\emaT \,,
\nn \\
\eqlab{yeye}
\eea
\eqlab{ydye}
\eseq
where
\bseq
\bea
|\alpha_{\Sd}| %= \dfrac{|a_{\Sd}|}{c_{\Sd}}
=% 8^{\frac 1 4}
\dfrac{2 m_s^{}}{m_b^{}}
\sqrt{\dfrac{m_d^{}}{\sqrt{2}m_s^{}}}\,, 
&\mbox{\quad}&
|\beta_{\Sd}|  %= \dfrac{|b_{\Sd}|}{c_{\Sd}}
= \dfrac{2 m_s^{}}{m_b^{}}\,,
\\
|\alpha_{\Se}| %= \dfrac{|a_{\Sd}|}{c_{\Sd}}
= %8^{\frac 1 4} \dfrac{\sqrt{m_e^{} m_\mu^{}}}{m_\tau^{}},
\dfrac{2 m_\mu^{}}{m_\tau^{}}
\sqrt{\dfrac{m_e^{}}{\sqrt{2}m_\mu^{}}}\,, 
&\mbox{\quad}&
|\beta_{\Se}| %= -3\dfrac{|a_{\Sd}b_{\Sd}|}{c_{\Sd}^2}
= \dfrac{2 m_\mu^{}}{m_\tau^{}}\,.
\eea
\eseq
 When we omit the first generation contribution by
setting $a_{\Su}=a_{\Sd}=0$,
the mixing angle between the second and the third generation
in the charged-lepton
mixing matrix $U_{\Se}$ of \eqref{UeUn} satisfies
\beq
\tan 2 \theta_{23}^e
\simeq \dfrac{2}{|\beta_{\Se}|^2-|\alpha_{\Se}|^2}
\simeq \dfrac{1}{2}\l(\dfrac{m_\tau}{m_\mu}\r)^2
\gg 1,
\eqlab{lepton23}
\eeq
where $\sin \theta^{\Se}_{23} = (U_{\Se})_{23}$.
 Therefor, if the neutrino-mixing matrix $U_{\nu}$
of \eqref{UeUn} is not too much different from the
diagonal matrix,
then desired boundary condition of $\sin^2 2\theta_{23} \simeq 1$
at the GUT scale can be obtained.
 This is in sharp contrast with the corresponding mixing angle
\beq
\tan 2 \theta^{\Sd}_{23}
\simeq \l|\beta_{\Sd}\r|
=2\dfrac{m_s^{}}{m_b^{}}
\ll 1\,,
\eqlab{quark23}
\eeq
in the down-quark sector.
 If the up-quark Yukawa matrix is approximately diagonal
(as in the Fritzsch form \eqref{up-1} with 
$a_{\Su} \ll b_{\Su} \ll c_{\Su}$),
the corresponding CKM matrix element remains small.
The two opposing results
follows essentially from the same Yukawa matrix in the
SU(5) theory.

 It has been well-known that the SU(5) constraint, \eqref{ydGUT},
on the lepton and down-quark Yukawa matrices leads to
the unacceptable mass relations:
$m_d/m_e = m_s/m_{\mu} = m_b/m_{\tau}$.
 A possible solution to this problem has been proposed by
Georgi and Jarlskog \cite{GJ}
where a ${\bf 45^{\ast}}$ Higgs boson gives
masses to the first two generation down quarks and 
charged-leptons\footnote{{
An alternative, but similar, solution in the supersymmetric SU(5)
model is found \eg in ref.\cite{SUSY45}.}}.
 By adopting their idea, we find that the lepton Yukawa matrix
of the form 
\beq
{\overline L_L} {\he} e_R^{} = {\overline L_L} 
\bmaT
   0    & a_{\Sd}e^{-i\phi_1} & 0      \\
a_{\Sd}e^{i\phi_1} &    0    & c_{\Sd}\\
   0    & -3b_{\Sd}e^{i\phi_2} & c_{\Sd}
\emaT
e_R^{}\,,
\eqlab{leptonMatrix}
\eeq
reproduces the desired mass formulas
\beq
m_e^{}   = \dfrac{v_{\Sd}}{\sqrt{2}}\dfrac{a_{\Sd}^2}{3b_{\Sd}},
\mbox{\hspace{5ex}}
m_\mu^{} = {v_{\Sd}}\dfrac{3b_{\Sd}}{2},
\mbox{\hspace{5ex}}
m_\tau^{}= {v_{\Sd}} c_{\Sd}.
\eqlab{me-1}
\eeq
 The Yukawa matrices, \eqref{down-1} for $\hd$
and \eqref{leptonMatrix} for $\he$,
can be obtained at the GUT scale if the element $\hd_{23}(\he_{32})$
is dominated by the coupling to the ${\bf 45^{\ast}}$ Higgs,
while the other elements come from the couplings to
the ${\bf 5^{\ast}}$ Higgs.
 The Yukawa matrices \eqref{up-1}, \eqref{down-1} and
\eqref{leptonMatrix} can \eg be obtained from the SU(5)
Lagrangian
\bea
{\cal L}_{yukawa}^{\rm GUT}
 &=& y^{10}_{i_{\Su} j_{\Su}} \varepsilon^{abcde}
                (\phi^{\Su})_a(\chi_{i_{\Su}})_{bc} \cdot (\chi_{j_{\Su}})_{de}
              + y^{5^{\ast}}_{i_{\Sd} j_{\Sd}}
               (\phi^{\Sd})^a(\chi_{i_{\Sd}})_{ab} \cdot (\psi_{j_{\Sd}})^b
\nn \\
           & & + y^{45^{\ast}}_{23}(\sigma)^{ab}_{c}
                   (\chi_2)_{ab} \cdot (\psi_3)^c
           + {\mbox{neutrino terms}} 
               + h.c.\,,
\eqlab{SU5-yukawa2}
\eea
where, $\sigma$ is the {\bf 45$^\ast$} Higgs particle.
All Yukawa elements are set to zero except
$(i_{\Su},j_{\Su})=(1,2)$, $(2,1)$, $(2,3)$, $(3,2)$, $(3,3)$,
and $(i_{\Sd},j_{\Sd})=(1,2)$, $(2,1)$, $(3,2)$, $(3,3)$.
It should be noted here that 
although the NNI texture of eqs.(\ref{eqn:up-1}), (\ref{eqn:down-1}),
and (\ref{eqn:leptonMatrix}) are achieved by using the weak-basis
transformation of the $\chi_i$ and $\psi_j$ fields at the GUT scale,
the dominance
of the ${\bf 45^{\ast}}$ coupling in the
$(i_{\Sd},j_{\Sd})=(2,3)$
coupling and the suppression of all the other 
${\bf 45^{\ast}}$ couplings are assumptions underlying
the Yukawa Lagrangian of \eqref{SU5-yukawa2}.

 Let us clarify our assumptions for the Yukawa matrices
at the GUT scale.
 It has been known that the zeros of the F-BS textures
are automatically obtained by making a proper 
weak-basis transformation,
because they are special cases of the generic NNI form \cite{NNI}.
 In the case of the SU(5) GUT, this statement needs
clarification because the left and right components of 
the up-quarks belong to the same representation and hence
they cannot be rotated independently.
 We find, however, that as long as we restrict
ourselves to the symmetric Yukawa matrix for the up-quark,
the Yukawa matrix $\hu$ can be transformed into the
Fritzsch form while retaining the general NNI form
for the down-quark Yukawa matrix.
 The proof is obtained by using the weak-basis transformation
of the $\chi({\bf 10})$ and $\psi({\bf 5^{\ast}})$ fields, 
similarly to that of Ref. \cite{Takasugi}.
 Thus, the textures of our Yukawa matrices $\hu$ and $\hd$($\he$)
are still general.
 It is the restriction of the down-quark Yukawa matrix
to the BS form in \eqref{down-1}
and the Georgi-Jarlskog modification for the charged-lepton
Yukawa matrix in \eqref{leptonMatrix}
 that consist our assumptions
\cite{no4}. 

 In addition, we have the Dirac-type Yukawa matrix for the neutrino ($\hn$)
and the Majorana type Yukawa matrix for the
right-handed neutrino (${\cal M_R}$) at the SU(5) GUT scale.
 We can transform ${\cal M_R}$ to the diagonal form in general,
but $\hn$ cannot be transformed to the NNI form at the same time.
 This is because in the SU(5) GUT, 
only three types of the matter fields, that transform as
${\bf 10}$, ${\bf 5^{\ast}}$, and ${\bf 1}$,
are rotated independently in the flavor space.
 The degrees of freedom associated with the weak-basis
transformation in the flavor space
can be used to make $\hu$
and $\hd$ in the NNI form and ${\cal M_{R}}$ diagonal,
but there is no freedom left to simplify the fourth matrix $\hn$.
 Therefore, {\it we need to assume the texture of $\hn$}
in order to make predictions on the MNS matrix elements.

 In summary, the matrices $\hu$ and $\hd(\he)$, 
are transformed to the F-BS form and ${\cal M_R}$ to the diagonal form
by using the weak-basis transformation degrees of freedom.
 However, the three eigenvalues of ${\cal M_R}$ and the texture of $\hn$
are not known.
 In the following section, 
we study consequences of a few simple textures of $\hn$
while taking ${\cal M_R}$ to be $M_R \times \id$,
where $M_R$ is the right-handed neutrino decoupling scale and
$\id$ is the $3 \times 3$ unit matrix.
 We note here that the light-neutrino mass matrix
$\kappa=\hn {\cal M_R}^{-1} \hn^T$ 
takes its most general form with ${\cal M_{R}}=M_R\times \id$,
as long as the matrix $\hn$ is taken general.
 It is worth noting here that when the original Majorana matrix
${\cal M_R}$ contains $CP$ violating phases,
the phases appear in the matrix $\hn$ in this basis.
 By taking $\hn$ to be real in the following analysis,
we assume $CP$ invariance in the $\nu_R$ mass matrix.
 This assumption affects our predictions for the
Majorana phases $\varphi_2$ and $\varphi_3$
of the MNS matrix elements at the weak scale.

\section{The texture of $\hn$ and the MNS matrix}
\label{sec:4}
\setcounter{equation}{0}
 In this section, we analyze the 
Maki-Nakagawa-Sakata (MNS) lepton-flavor mixing matrix
by using the 1-loop RGE of the MSSM from the GUT scale
($M_{GUT} = 1.7 \times 10^{16}$ GeV) to the $m_Z^{}$ scale.
 
 We report the results of our exploratory studies where
we examine consequences of a few simple textures of the matrix
$\hn$. We first examine the case of the simplest texture,
the diagonal $\hn$,
and study consequences of the two choices of the right-handed neutrino
decoupling scale,
$M_R=M_{GUT}$ and $M_R=3\times 10^{14}$ GeV.
In the latter part, we study consequences
of the Fritzsch-type texture.

\subsection{Diagonal $\hn$ texture}
We first study consequences of diagonal $\hn$.
The Yukawa matrices take the following forms at the 
SU(5) GUT scale
\bea
\hu =
\bmaT
   0    & a_{\Su} & 0      \\
a_{\Su} &    0    & b_{\Su}\\
   0    & b_{\Su} & c_{\Su}
\emaT\,,
&\mbox{\quad\quad} &
\hn = 
\bmaT
a_\nu &   0   &    0    \\
  0   & b_\nu &    0    \\
  0   &   0   & c_\nu  
\emaT\,,
\nn \\
\hd = 
\bmaT
   0                & a_{\Sd} e^{i\phi_1}& 0      \\
a_{\Sd}e^{-i\phi_1} &    0               & b_{\Sd} e^{i\phi_2}\\
   0                & c_{\Sd}            & c_{\Sd}
\emaT\,,
&\mbox{\quad\quad} &
\he = 
\bmaT
   0    & a_{\Sd}e^{-i\phi_1}   & 0      \\
a_{\Sd}e^{i\phi_1} &    0      & c_{\Sd}\\
   0    & -3b_{\Sd}e^{i\phi_2} & c_{\Sd}
\emaT\,.
\eqlab{simple1}
\eea
 Here we assume for brevity that all elements of $\hn$,
$a_{\nu}$, $b_{\nu}$, and $c_{\nu}$ are real.

 We have 11 parameters in the Yukawa matrices.
 On the other hand, there are 22 observables:
12 fermion masses, 
3 angles and 1 phase each for the CKM matrix and 
the MNS matrix, 
and the 2 more Majorana phases of the MNS matrix.
Thus, there are 11 predictions,
whereas two Majorana phases $\varphi_2$ and $\varphi_3$
are unobservable in the neutrino oscillation
experiments\footnote{They are observable
in the lepton-number violating processes
\eg in the neutrino-less double beta decays.}.
The generation hierarchy among the $\hn$ elements,
\bea
a_{\nu} \ll b_{\nu} \ll c_{\nu}\,,
\eqlab{hie_para}
\eea
follows from our assumption of $m_1 \ll m_2 \ll m_3$,
while the following relations
\beq
a_{\Su} \ll b_{\Su} \ll c_{\Su}\,,
\mbox{\quad}
a_{\Sd} \ll b_{\Sd} \ll c_{\Sd}\,,
\eeq
are needed to reproduce the known quark and lepton masses.

 We set $\tan \beta = 3$ in our numerical evaluation of the 
1-loop RGE's.
The gauge couplings are set at $\alpha_1(m_Z^{})=0.017$ and
$\alpha_2(m_Z^{})=0.034$.
 If $\tan \beta$ is significantly smaller than 2, the Yukawa coupling of the
top quark blows up below the GUT scale.
 The fits with the experimental values of the CKM matrix elements
become worse as $\tan \beta$ increases \cite{no4}.

 The parameters $a_{\Su}$, $b_{\Su}$, $c_{\Su}$
and $a_{\Sd}$, $b_{\Sd}$, $c_{\Sd}$
are fixed, respectively, by the central values of the
up-type-quark and the charged-lepton masses, 
as shown in \eqref{up-part} and \eqref{charged-lepton}.
 $a_{\nu}$, $b_{\nu}$ and $c_{\nu}$ are fixed by 
the neutrino masses which are 
chosen in the allowed range of \Tbref{MNS-Mass}.
Phases in the $\hd$ ($\he$) are constrained by the CKM matrix
elements $|V_{us}|$ and $|V_{cb}|$.
The three down-type quark masses, the CKM matrix element $|V_{ub}|$,
$J_{\rm CKM}^{}$, and all the six independent parameters
of the MNS matrix can be predicted.

 By numerical analysis of the RGE, we obtain the down-type quark masses
\bea
m_b(m_Z^{}) &=& 3.3   {\mbox {\hspace{6ex} GeV}}\,, \nn \\
m_s(m_Z^{}) &=& 0.081 {\mbox {\hspace{4ex} GeV}}\,, \nn \\
m_d(m_Z^{}) &=& 0.0032 {\mbox {\hspace{3ex} GeV}}\,. 
\eea
 These values are roughly consistent with the experimental values 
of \eqref{down-part},
reconfirming the validity of the Georgi-Jarlskog scenario \cite{GJ}.
 The CKM matrix elements are
\bea
V_{us} = 0.22\,,
\mbox{\quad}
V_{cb} = 0.046\,,
\mbox{\quad}
\dfrac{\l|V_{ub}\r|}{V_{cb}} = 0.10\,,
\nn \\
J_{\rm CKM}^{} = 3.6 \times 10^{-5}~~
\l(\delta_{\rm CKM} = \degree{78.1}\r)\,,
\eqlab{CKM_Simple}
\eea
when we take the parameters in \Tbref{other-one}.
\begin{table}[htbp]
\begin{center}
\begin{tabular}[tb]{|c|c|c|c|c|c|c|c|}
\hline
$a_{\Su}$ & $b_{\Su}$ & $c_{\Su}$ & $a_{\Sd}$ & $b_{\Sd}$ & $c_{\Sd}$ &
$\phi_1$ & $\phi_2$  \\
\hline
\hline
$\numt{1.1}{-4}$ &  $\numt{4.0}{-2}$ & 0.93 &
$\numt{1.0}{-4}$ & $\numt{5.9}{-4}$ & $\numt{1.5}{-2}$ &
$60^{\circ}$ &  $253^{\circ}$    \\
\hline
\end{tabular}
\caption{The input parameters at the GUT scale.}
\tblab{other-one}
\end{center}
\end{table}
 They are also consistent with 
the experimental constraint of \eqref{exp_ckm}.
 The results on the quark masses and the CKM matrix elements are
almost common in all the analyses below.
\subsubsection{$M_R=M_{GUT}$}
 We first study the simplest case of $M_R=M_{GUT}$,
where the light neutrino Yukawa matrix $\kappa$
takes the following form
\beq
\kappa = \hn {\cal M_R}^{-1} \hn^{T} =
\dfrac{1}{M_{GUT}}
\bmaT
a_\nu^2 &    0    &   0   \\
  0     & b_\nu^2 &   0   \\
  0     &    0    &  c_\nu^2   
\emaT\,,
\eeq
at the GUT scale.
 When the neutrino mass $m_2$ is fitted
to the MSW small mixing solution of \eqref{mass_2_MSWS},
$(m_1,m_2,m_3)=(0.0003,0.003,0.03)$eV for definiteness,
the MNS matrix elements become
\beq
  U_{e2} = 0.058\,, \mbox{\quad}
  |U_{e3}| = 0.058\,, \mbox{\quad}
  U_{\mu 3} = 0.70\,. 
\eeq
 These values are consistent with the corresponding 
experimental constraints
that are summarized in \Tbref{MNS-Mass}.
 The values of $a_{\nu}$, $b_{\nu}$ and $c_{\nu}$
at the GUT scale are
\beq
a_{\nu} = 0.54\,,
\quad
b_{\nu} = 1.7\,,
\quad
c_{\nu} = 5.4\,.
\eeq
The results are summarized in \Tbref{MGUT-diag}
in the first row (MSW-S).
\begin{table}[htbp]
\begin{center}
\begin{tabular}[tb]{|l||c|c|c||c|c|c|c|c|c|}
\hline
     &$a_{\nu}$ & $b_{\nu}$ & $c_{\nu}$ &
      $m_1$(eV) & $m_2$(eV) & $m_3$(eV)&
 $U_{e2}$ & $|U_{e3}|$ & $U_{\mu 3}$ \\
\hline
\hline
MSW-S&0.54& 1.7& 5.4&
 0.0003 & 0.003 & 0.03 &
  0.058 & 0.058 & 0.70  \\
\hline
MSW-L&0.70 & 2.2 & 4.4 &
 0.0005 & 0.005 & 0.02 &
 0.058 & 0.058 & 0.70 \\
\hline
V-O  &0.031&0.10&4.4&
 $\numt{1}{-6}$ & $\numt{1}{-5}$ & 0.02 &
  0.058 & 0.058 & 0.70 \\
\hline
\end{tabular}
\caption{The input parameters $(a_\nu,b_\nu,c_\nu)$ and the predictions when we take $\hn$ diagonal and $M_R=M_{GUT}$.}
\tblab{MGUT-diag}
\end{center}
\end{table}

 The results for the MSW large-mixing solution (MSW-L) and 
the vacuum-oscillation solution (V-O) are also shown in \Tbref{MGUT-diag}.
Because we use the diagonal $\hn$ texture,
the MNS matrix elements are essentially determined by the
matrix $\he\he^{\dagger}$.
The large $U_{\mu 3}$ and small $U_{e2}$ and $|U_{e3}|$
then follow almost independently of the
input neutrino mass values.
 Summing up, with the diagonal $\hn$ texture,
we can reproduce the MSW-S solution but not the other two solutions.

 The $CP$ violating parameters are listed in the \Tbref{MGUT-diag2}.
\begin{table}[htbp]
\begin{center}
\begin{tabular}[tb]{|l||c|c|c|}
\hline
     & $J_{\rm MNS}^{}$ &$\varphi_2^{}$ & $\varphi_3^{}$ \\
\hline
\hline
MSW-S & $-3.4 \times 10^{-11}$ &$\degree{133.0}$&$\numt{-7.8}{-9}$\\
\hline
MSW-L & $-8.5 \times 10^{-11}$ &$\degree{133.0}$&$\numt{-2.0}{-8}$ \\
\hline
V-O   & $-1.2 \times 10^{-13}$ &$\degree{133.0}$&$\numt{-3.9}{-11}$\\
\hline
\end{tabular}
\caption{The predicted values of the $CP$ violating parameters of the MNS matrix,
 $J_{\rm MNS}$, $\varphi_2^{}$ and $\varphi_3^{}$ when we take $\hn$ diagonal and $M_R=M_{GUT}$.}
\tblab{MGUT-diag2}
\end{center}
\end{table}
 The magnitude of the $CP$ violation parameter $J_{\rm MNS}^{}$
remains small when $M_R=M_{\rm GUT}$.
 The remaining two angles,
$\varphi_2$ and $\varphi_3$ in the MNS matrix 
can also be predicted\footnote{{
See Appendix C for more details.}}.
 Because we neglect the Majorana phases of the $\nu_R$
mass matrix ${\cal M_R}$, by taking $\hn$ to be real,
they are determined essentially by the unitary matrix $U_e$ that
diagonalize $\he\he^{\dagger}$; see eqs (C.7) and (C.8)
in Appendix C.
  The magnitudes of $\varphi_2^{}$ for all solutions are
large and the magnitudes of $\varphi_3^{}$ for all solutions
are small, reflecting the phases structure of \eqref{yeye}
at the GUT scale.

\subsubsection{$M_R=3\times10^{14}$ GeV}
When $M_{R}$ is lower than the $M_{GUT}$,
the RGE's including the Yukawa matrix $\hn$,
\eqref{RGE-high} of Appendix B,
apply in the region $M_R<\mu<M_{GUT}$,
while the terms proportional to $\hn$ decouple below $\mu=M_R$,
and the RGE of $\kappa$, \eqref{RGE_kappa}, takes over.
The magnitudes of the MNS matrix elements are only slightly
different from the $M_R=M_{GUT}$ case:
\beq
U_{e2} = 0.057\,, \mbox{\quad}
|U_{e3}| = 0.058\,, \mbox{\quad}
U_{\mu 3} = 0.71\,,
\eeq
when $m_2$ is fitted to the MSW small mixing solution.
 No improvements are found for the other two scenarios;
the elements $|U_{e2}|$ and $|U_{e3}|$ remain too small for the large mixing
solutions.
 The magnitudes of the input parameters at the GUT scale change 
significantly:
\beq
a_{\nu} = 0.071\,,
\quad
b_{\nu} = 0.23\,,
\quad
c_{\nu} = 0.74\,.
\eeq
$c_\nu$ and $c_{\Su}$ (see \Tbref{other-one}) are now comparable in magnitude.
The results are summarized in \Tbref{MR-diag}.
\begin{table}[htbp]
\begin{center}
\begin{tabular}[tb]{|l||c|c|c||c|c|c|c|c|c|}
\hline
& $a_{\nu}$ & $b_{\nu}$ & $c_{\nu}$&
  $m_1$(eV) & $m_2$(eV) & $m_3$(eV)&
  $U_{e2}$ & $|U_{e3}|$ & $U_{\mu 3}$ \\
\hline
\hline
MSW-S&0.071&0.23& 0.74&
 0.0003 & 0.003 & 0.03 &
 0.057 & 0.058 & 0.71  \\
\hline
MSW-L&0.093 & 0.30 & 0.60&
 0.0005 & 0.005 & 0.02 &
 0.057 & 0.058 & 0.71 \\
\hline
V-O  &0.0043&0.013&0.60&
  $\numt{1}{-6}$ & $\numt{1}{-5}$ & 0.02 &
 0.057 & 0.058 & 0.71  \\
\hline
\end{tabular}
\caption{The input parameters ($a_\nu$, $b_\nu$, $c_\nu$) and the
 predictions when we take $\hn$ diagonal and $M_R=3\times 10^{14}$ GeV.}
\tblab{MR-diag}
\end{center}
\end{table}

 The predictions for the $CP$ violating parameters are listed in
\Tbref{CP1}.
\begin{table}[htbp]
\begin{center}
\begin{tabular}[tb]{|l||c|c|c|}
\hline
 & $J_{\rm MNS}^{}$ & $\varphi_2^{}$ & $\varphi_3^{}$ \\
\hline
\hline
MSW-S 
& $\numt{-8.6}{-8}$ & $\degree{133.0}$ & $\numt{\degree{-1.5}}{-3}$\\
\hline
MSW-L 
 & $\numt{-6.0}{-8}$ & $\degree{133.0}$ & $\numt{\degree{-3.1}}{-3}$\\
\hline
V-O   
 & $-1.1 \times 10^{-6}$ & $\degree{133.2}$ & $\numt{\degree{-9.9}}{-5}$\\
\hline
\end{tabular}
\caption{Predictions for the $CP$ violating parameters
of the MNS matrix, $J_{\rm MNS}$, $\varphi_2$, $\varphi_3$,
for diagonal $\hn$ and $M_R=3\times 10^{14}$ GeV.}
\tblab{CP1}
\end{center}
\end{table}
 The magnitude of the parameter $J_{\rm MNS}^{}$ is now found to be much
bigger than the $M_R=M_{GUT}$ case shown in \Tbref{CP1}.
 This is because the matrix $\hn$ acquires phases
by the RGE effects at $M_R < \mu < M_{GUT}$.
 The magnitude of the one of the Majorana phases
$\varphi_3^{}$ is also found to be bigger than the $M_R=M_{GUT}$
case, reflecting the renormalization effect on $\he$
at $M_R < \mu < M_{GUT}$.
 On the other hand, 
the magnitude of $\varphi_2^{}$ is not sensitive to the $M_R$,
because the phase, $\varphi_2^{}$, is essentially
determined by $\arg (U_{e2})$.

\subsection{Fritzsch-type $\hn$ texture}
In this subsection, we study consequence of the Fritzsch form $\hn$.
The Yukawa matrix of $\hn$ takes the following form
at the SU(5) GUT scale,
\beq
\hn =
\bmaT
  0   & a_\nu &   0   \\
a_\nu &   0   & b_\nu \\
  0   & b_\nu & c_\nu
\emaT\,,
\eqlab{c2}
\eeq
where all elements of $\hn$, $a_\nu^{}$, $b_\nu^{}$, and $c_\nu^{}$
are real for brevity.
 We set the right-handed neutrino decoupling scale,
$M_R$, to be $\numt{3}{14}$ GeV.

 Here also we have 11 parameters in the Yukawa matrices and
hence 11 predictions.
 The 8 parameters of the Yukawa matrices $\hu$ and $\hd$ are
fixed by the three up-quark masses, the three charged-lepton masses,
and the two CKM matrix elements,
$V_{us}$ and $V_{cb}$, as in the previous case.
 The magnitudes of the input values, $a_{\Su}$, $b_{\Su}$, $c_{\Su}$,
$a_{\Sd}$, $b_{\Sd}$, $c_{\Sd}$, $\phi_1$ and $\phi_2$
hence take the same values as
those listed in \Tbref{other-one}.
 The predictions for the three down-type quark masses and
the remaining CKM matrix elements are also the same.

 On the other hand,
the magnitudes of the input parameters, $a_{\nu}^{}$,
$b_{\nu}^{}$, and $c_{\nu}^{}$ at the GUT scale 
should be bigger than those of the diagonal $\hn$ case
in order to obtain the same neutrino masses.
 For instance, $c_{\nu}^{}$ is now bigger than $c_{\Su}$
see \Tbref{c2t} and \Tbref{other-one}.
\begin{table}[htbp]
\begin{center}
\begin{tabular}[tb]{|l||c|c|c||c|c|c|c|c|c|}
\hline
      &$a_{\nu}$&$b_{\nu}$&$c_{\nu}$
      &$m_1$(eV) &$m_2$(eV)&$m_3$(eV) 
      &$U_{e2}$&$|U_{e3}|$&$U_{\mu 3}$ \\
\hline
\hline
MSW-S & 0.01 & 0.49 & 1.1 
      & $\numt{1.3}{-8}$ & 0.002& 0.08
      & 0.054 & 0.033 & 0.42  \\
\hline
MSW-L & 0.13 & 0.44 & 1.1 
      & $\numt{3.0}{-4}$ & 0.003 & 0.08
      & 0.45 & 0.029 & 0.45  \\
\hline
V-O   & 0.0065 & 0.10 & 1.28
      & $\numt{8}{-7}$ & $\numt{8}{-6}$ & 0.08
      & 0.45 & 0.054 & 0.66  \\
\hline
\end{tabular}
\caption{The input input parameters ($a_\nu$, $b_\nu$, $c_\nu$)
and the predictions
when we take the Fritzsch-type $\hn$ (at the GUT scale)
and $M_R=\numt{3}{14}$GeV.}
\tblab{c2t}
\end{center}
\end{table}
 By comparing \Tbref{c2t} and \Tbref{MR-diag}, we find
that by choosing the Fritzsch form the ``hierarchy''
between $b_\nu$ and $c_\nu$ is very weak,
$b_\nu/c_\nu \simeq 0.4$, for the MSW solutions.
 This has significant consequences in the predictions
for the MNS matrix elements:
\bseq
\bea
& &
U_{e2} = 0.054\,,
{\mbox{\quad}}
|U_{e3}| = 0.033\,,
{\mbox{\quad}}
U_{\mu 3} = 0.42\,,
{\mbox{\quad}}
\mbox{({\sf MSW-S})}\,, \\ 
& &
U_{e2} = 0.45\,,
{\mbox{\quad}~~}
|U_{e3}| = 0.029\,,
{\mbox{\quad}}
U_{\mu 3} = 0.45\,,
{\mbox{\quad}}
\mbox{({\sf MSW-L})}\,, \\ 
& &
U_{e2} = 0.45\,,
{\mbox{\quad}~~}
|U_{e3}| = 0.054\,,
{\mbox{\quad}}
U_{\mu 3} = 0.66\,,
{\mbox{\quad}}
\mbox{({\sf V-O})}\,.  
\eea
\eseq
 We can now reproduce vacuum-oscillation solution with
the Fritzsch form $\hn$.
 However, the predictions for the $U_{\mu 3}$ element
are slightly smaller than the experimental
constraint of \Tbref{MNS-Mass} for the MSW solutions.
 In the diagonal $\hn$ case, the MNS matrix elements are essentially
determined by the matrix $\he \he^{\dagger}$,
and large $U_{\mu 3}$ results as a consequence of 
large $\tan 2\theta_{23}^e$ in \eqref{lepton23}. 
 The prediction holds with the Fritzsch form for the V-O solution
because the hierarchy $b_{\nu} \ll c_{\nu}$ in the third row of 
\Tbref{c2t} implies essentially diagonal $\hn\hn^{T}$.
 The predicted values of the $U_{\mu 3}$ element reduces significantly
for the MSW solutions because the absence of the hierarchy,
${b_\nu}/{c_\nu} \simeq 0.4$, implies significantly non-diagonal
$\hn \hn^{T}$.
 By choosing large $m_3$ and small $m_2$ within the allowed ranges
of \Tbref{MNS-Mass}, $U_{\mu 3}$ can increase up to 0.42 for MSW-S and
0.45 for MSW-L solutions,
slightly below the range allowed by 
Super-Kamiokande in \Tbref{MNS-Mass}.

 The matrix element $|U_{e3}|$ remains small in all
three solutions.
 The element $U_{e2}$ can also be made barely consistent
with the allowed ranges of \Tbref{MNS-Mass} for all the
solutions.
 Because the $U_{e2}$ element from the diagonalization
of $\he\he^{\dagger}$ is rather small
($U_{e2}=0.057$ in \Tbref{MR-diag} when $\hn$ is diagonal)
large $U_{e2}$ results from the diagonalization of $\hn\hn^{T}$.
 We find that $U_{e2}$ is proportional to $\sqrt{m_1/m_2}$,
and $U_{e2}=0.45$ in \Tbref{c2t} results when $m_1/m_2=1/10$.
 In order to accommodate small $U_{e2}$ in the MSW-S solution,
$m_1/m_2$ should be chosen small.
 The minimal of ${U_{e2}}$ is found to be $U_{e2}=0.054$ at
$m_1/m_2=\numt{7}{-6}$.

 The magnitude of the parameter $J_{\rm MNS}$ in \Tbref{c2t2}
is now found to be much bigger than that of the diagonal
$\hn$ case in \Tbref{CP1} for all three cases.
\begin{table}[htbp]
\begin{center}
\begin{tabular}[tb]{|l||c|c|c|}
\hline
&$J_{\rm MNS}^{}$ & $\varphi_2^{}$ & $\varphi_3^{}$ \\
\hline
\hline
MSW-S & $\numt{-4.3}{-4}$ & $\degree{89.7}$ & $\degree{0.014}$\\
\hline
MSW-L & $\numt{-4.2}{-3}$ & $\degree{5.8}$ & $\degree{0.26}$\\
\hline
V-O   & $\numt{-8.4}{-3}$ & $\degree{5.0}$ & $\degree{-0.057}$\\
\hline
\end{tabular}
\caption{Predictions for the $CP$ violating parameters
of the MNS matrix, $J_{\rm MNS}$, $\varphi_2$, $\varphi_3$,
for the Fritzsch-type $\hn$ (at the GUT scale) and $M_R=\numt{3}{14}$GeV.}
\tblab{c2t2}
\end{center}
\end{table}
 Predicted magnitudes of $J_{\rm MNS}$ are now bigger than that
of $J_{\rm CKM}$ in \eqref{CKM_Simple}.
 This is because the only non-real element of $\he \he^{\dagger}$
in \eqref{yeye} is in the (1,3) element and has small magnitude.
 Large $|J_{\rm MNS}|$ results only with significantly non-diagonal $\hn$.
 We also note that the magnitude of the Majorana phase
$\varphi_3$ is now larger than 
that of the diagonal $\hn$ case and the sign of $\varphi_3^{}$
is changed in the MSW solutions.
 Their magnitudes remain smaller than $\degree{1}$, however.
 We find significantly different predictions for $\varphi_2$
as compared to the diagonal $\hn$ case;
$\varphi_2 \simeq \degree{133}$ in \Tbref{CP1}.
 Significant fraction of the
contribution from $\he \he^{\dagger}$
is canceled by the non-diagonality of $\hn$.
 The magnitude of $\varphi_2^{}$ in the
``MSW-S'' case is large ($\varphi_2 \simeq \degree{90}$ in \Tbref{c2t2})
because of our choice of very small
$a_{\nu}^{}$ to accommodate small $U_{e2}$.
 If we take $a_{\nu}^{}=0.11$ for $m_1=m_2/10$,
the predictions are
\beq
U_{e2} = 0.45\,,
\mbox{\quad}
|U_{e3}| = 0.027\,,
\mbox{\quad}
U_{\mu 3}=0.48\,,
\eqlab{111}
\eeq
and
\beq
J_{\rm MNS} = -\numt{4.8}{-3}\,,
\mbox{\quad}
\varphi_2^{} = \degree{5.7}\,,
\mbox{\quad}
\varphi_3^{} = \degree{0.18}\,.
\eqlab{222}
\eeq
 Therefore large $U_{e2}$ and $U_{\mu 3}$, large $|J_{\rm MNS}|$
and small $\varphi_2$, $\varphi_3$ are natural consequences of the
Fritzsch type $\hn$ in our model.
 In order to accommodate small $U_{e2}$ for the MSW small-mixing 
solution, we should assume $m_1/m_2 \simeq \numt{7}{-6}$.
 Note, however, that our predictions for the Majorana
phases $\varphi_2$ and $\varphi_3$ assume $CP$-invariance in 
the right-handed neutrino mass matrix ${\cal M_R}$.

Finally, we report our finding for $\hn=\hu$, 
\ie when we impose the following conditions
\beq
a_{\Su} = a_{\nu}\,,
\mbox{\quad}
b_{\Su} = b_{\nu}\,,
\mbox{\quad}
c_{\Su} = c_{\nu}\,,
\eeq
at the GUT scale.
 In this case, we have three more predictions.
 In particular, the neutrino masses can be predicted
for a given value of the right-handed-neutrino decoupling scale $M_R$. 
 By choosing $M_R=\numt{1.7}{14}$GeV, we find
\bea
m_1 &=& \numt{4.8}{-12}{\mbox { eV}}\,, \nn \\
m_2 &=& \numt{3.4}{-7}{\mbox {~~eV}}\,, \nn \\
m_3 &=& 0.08{\mbox {~~eV}}\,.
\eea
The mass squared differences are
\bea
O(m_3^2-m_2^2) \simeq \numt{6}{-3} \mbox{eV$^2$}\,, \nn \\
O(m_2^2-m_1^2) \simeq \numt{1}{-13} \mbox{eV$^2$}\,.
\eea
The difference $m_2^2-m_1^2$ is too small even for the V-O
solution of the solar-neutrino experiments.

\section{Summary}
\label{sec:sum}
\setcounter{equation}{0}
 In this article,
we explain the imbalance of the flavor-mixing-angles
between the quark and the lepton sectors
suggested by the recent neutrino-oscillation experiments
\cite{CHOOZ}-\cite{SK1} in the 
supersymmetric SU(5) GUT with the see-saw mechanism.
 Especially, we look for the reason why the matrix element
$|U_{\mu 3}|$ is much larger than the corresponding 
$|V_{cb}|$ of the CKM matrix.
 
 We use two tools for analyzing the cause
of the imbalance between the CKM matrix and the 
corresponding MNS matrix for lepton-flavor mixing \cite{MNS,CKM}.
 One is the RGE of the MSSM
between the GUT scale and the weak scale.
 The RGE of the neutrino-mass matrix \cite{nue-RGE}
generated by the see-saw mechanism
implies that a large mixing between the second and third generation
leptons is obtained at the $m_Z^{}$ scale as long as the mixing
is large at the GUT scale.
 The other is the SU(5) constraints on the Yukawa matrices
at the GUT scale.
 In particular, $\hu=\hu^T$ and $\hd=\he^T$ follow if the Higgs
 doublets of the MSSM belong to ${\bf 5}$ and ${\bf 5^{\ast}}$
representations of SU(5) \cite{up2now2}. 
 By choosing an appropriate weak basis
at the GUT scale,
we can take $\hu$ to have the symmetric Fritzsch form \cite{fritzsch}
while $\hd$ should take the generic NNI form \cite{Takasugi}.
 The observed quark masses and the CKM matrix elements
then forces $\hd$ to take the asymmetric BS form \cite{BS}
where the largest elements appear in the bottom row
of the second and the third column \cite{Takasugi2}.
 The SU(5) relation then forces $\he$ to have the
largest elements in the third column at the second and the third row.
 We show that this asymmetry between $\hd$ and $\he$ leads
to the asymmetry in the unitary matrices that diagonalize
the down-quark and the charged-lepton mass matrices.
 In particular, $|V_{cb}| \ll 1$ and $|U_{\mu 3}| \sim 0.7$
can be obtained at the same time
if the neutrino mass matrix is approximately diagonal.
 We present a few examples by adopting simple textures for the
neutrino Yukawa matrix $\hn$,
in the basis where the heavy right-handed-neutrino
Majorana mass matrix ${\cal M_R}$ is diagonal.

 The favorable MNS matrix elements are thus obtained together with the
acceptable $m_{\tau}/m_b$ ratio from the simplest SU(5) Yukawa sector.
As has been well-known, however,
this model gives unacceptable predictions
for the other ratios, $m_{\mu}/m_s$ and $m_e/m_d$.
 In our actual numerical calculation,
we modify the SU(5) Yukawa sector by introducing a ${\bf 45^{\ast}}$
Higgs boson coupling between the second generation decuplet (${\bf 10}$)
and the third generation quintet (${\bf 5^{\ast}}$),
in order to accommodate the 
Georgi-Jarlskog mass relations \cite{GJ} $m_{\mu}/m_s = m_d/m_e = 3$
at the GUT scale.
 Acceptable quark and lepton masses then follow for not too large 
$\tan \beta$ \cite{no4}.
 The mixing between the second and third generation leptons,
$|U_{\mu 3}|$, is insensitive to this modification,
while the others, $|U_{e2}|$ and $|U_{e3}|$,
are affected significantly.

 We analyze the Maki-Nakagawa-Sakata (MNS) \cite{MNS} lepton-flavor
mixing matrix and neutrino masses numerically by using
the 1-loop RGE of the MSSM.
 First, we study the case where the texture of $\hn$
is diagonal,
and examine the sensitivity of the results to 
the right-handed-neutrino decoupling scale $M_R$, by setting
$M_R=M_{GUT}$ or $M_R=\numt{3}{14}$GeV.
 We can reproduce the MSW small-mixing solution 
but not the other solutions of the solar-neutrino deficit
with the diagonal $\hn$, almost independent of $M_R$.
 The magnitudes of the $CP$ violating parameters of the MNS matrix,
$J_{\rm MNS}$ and $\varphi_3^{}$ are very small,
while large $\varphi_2$
can arise even when the heavy Majorana mass matrix
${\cal M_R}$ does not have a $CP$ violating phase. 
 We also study the case where $\hn$ has the texture of 
Fritzsch form, by setting $M_R=\numt{3}{14}$GeV.
 We find the tendency to have both $|U_{\mu 3}|$ and $|U_{e2}|$
large when $m_1/m_2\simeq1/10$.
 MSW large-angle solution and the vacuum-oscillation solution
can hence be reproduced.
 In order to accommodate the MSW small-angle solution,
we need to fine-tune the mass ratio to be $m_1/m_2 \simeq \numt{7}{-6}$. 
 The magnitude of the $CP$ violating parameter $J_{\rm MNS}$
can now be bigger than that of $J_{\em CKM}$,
while the Majorana phases $\varphi_2$ and $\varphi_3$ are small
for the large angle solutions.
 As a special case, we examine the consequence of $\hu=\hn$
at the GUT scale and find that the ratios of the three neutrino masses
are inconsistent with the observation.

 We can naturally explain the imbalance of the mixing angles
between the lepton and quark sectors within the SU(5) GUT.
 We hope that our finding may shed light on the Yukawa sector of the
supersymmetric GUT theories.

\section*{Acknowledgment}
 The authors thanks M.~Tanimoto, S.~Alam, G.-C.~Cho, and Y.~Okada
for useful discussion and comments.
 N.O. wishes to thank N.~Haba for his useful discussion
and warm encouragement.
 This work is supported in part by Grant-in-Aid
for Scientific Research on Priority Areas (A).
 The work of N.O. is supported by the JSPS Research Fellowships
for Young Scientists, \#2996.

\appendix
\label{app:0}
\setcounter{equation}{0}
\section{Vacuum oscillation probabilities and the MNS matrix.}
The transition probability $P_{\nu_\alpha \to \nu_\beta}$,
 $(\alpha \neq \beta)$ is
\bea
P_{\nu_\alpha \to \nu_\beta} &=&
\l|\sum_{j=1}^3 \l(V_{\rm MNS}^{}\r)_{\beta j}^{} 
\exp\l({\dfrac{-i m_j^2}{2E}L}\r)
\l(V_{\rm MNS}^{\dagger}\r)_{j \alpha}^{} \r|^2 \nn \\
 &=& 
\l| U_{\beta 1}^{} U_{\alpha 1}^{\ast}
+U_{\beta 2}^{}
  \exp\l({\dfrac{-i \delta m_{12}^2}{2E}L}\r)
  U_{\alpha 2}^{\ast}
+U_{\beta 3}^{}
  \exp\l({\dfrac{-i \delta m_{13}^2}{2E}L}\r)
  U_{\alpha 3}^{\ast}
\r|^2,
\eqlab{pro1}
\eea
where $\delta m_{ij}^2 = m_j^2 - m_i^2$,
and we used the identity
\beq
\l(V_{\rm MNS}^{}\r)_{\beta j}^{}
\l(V_{\rm MNS}^{\dagger}\r)_{j \alpha}^{}
=
U_{\beta j}^{} U^{\ast}_{\alpha j}
\eeq
which follows directly from our parameterization \eqref{MNS1} of
the MNS matrix.
It reduces to simple forms when the conditions
\beq
\delta m_{12}^2 \ll m_{13}^2
\eeq
holds.
 there are two cases.

When the conditions 
\beq
\dfrac{\delta m_{12}^2}{2E}L \ll 1
\sim \dfrac{\delta m_{13}^2}{2E}L
\eqlab{A-massd}
\eeq
are satisfied, \eqref{pro1}
can be simplified significantly:
\bea
P_{\nu_\alpha \to \nu_\beta} &=&
\l|-U_{\beta 3}^{} U_{\alpha 3}^{\ast}
+U_{\beta 3}^{}\exp\l({\dfrac{-i \delta m_{13}^2}{2E}L}\r) U_{\alpha 3}^{\ast}
\r|^2 \nn \\
&=&
4\l|U_{\alpha 3}^{}\r|^2\l|U_{\beta 3}^{}\r|^2 \sin^2
\l(
{\dfrac{\delta m_{13}^2}{4E}L}
\r).
\eqlab{pro2}
\eea
On the other hand when
\beq
\dfrac{\delta m_{12}^2}{2E}L \sim 1 
\ll \dfrac{\delta m_{13}^2}{2E}L
\label{ap-cond-2}
\eeq
holds,
the probability \eqref{pro1} takes the following form:
\beq
P_{\nu_{\alpha} \to \nu_{\beta}}
=
2|U_{\alpha 3}^{}|^2|U_{\beta 3}^{}|^2
-\l[
4 Re(U_{\alpha 1}^{}U_{\beta 1}^{\ast}U_{\beta 2}^{}U_{\alpha 2}^{\ast})
\sin^2\l(\dfrac{\delta m_{12}^2}{4E}L\r)
+2 J_{\rm MNS}^{}
\sin^2\l(\dfrac{\delta m_{12}^2}{2E}L\r)
\r]\,,
\eeq
where $J_{\rm MNS}^{}=Im(U_{\alpha 1}^{}U_{\beta 1}^{\ast}
 U_{\beta 2}^{} U_{\alpha 2}^{\ast}$).

The survival probability is,
\bea
P_{\nu_\alpha \to \nu_\alpha} &=&
\l|\sum_{j=1}^3 \l(V_{\rm MNS}^{}\r)_{\alpha j}^{}
 \exp\l({\dfrac{-i m_j^2}{2E}L}\r)
\l(V_{\rm MNS}^{\dagger}\r)_{j \alpha}^{}\r|^2 \nn \\
 &=& 
\l|U_{\alpha 1}^{} U_{\alpha 1}^{\ast}
+U_{\alpha 2}^{}
  \exp\l({\dfrac{-i \delta m_{12}^2}{2E}L}\r)
  U_{\alpha 2}^{\ast}
+U_{\alpha 3}^{}
  \exp\l({\dfrac{-i \delta m_{13}^2}{2E}L}\r)
  U_{\alpha 3}^{\ast}
\r|^2.
\eea
Under the condition (\ref{eqn:A-massd}), we find
\bea
P_{\nu_\alpha \to \nu_\alpha} &=&
\l|1-U_{\alpha 3}^{} U_{\alpha 3}^{\ast}
 +U_{\alpha 3}^{}
  \exp\l({\dfrac{-i \delta m_{13}^2}{2E}L}\r)
  U_{\alpha 3}^{\ast}
\r|^2 \nn \\
&=&
 1 - 4\l|U_{\alpha 3}^{}\r|^2 \l(1-\l|U_{\alpha 3}^{}\r|^2 \r)
\sin^2
\l(
{\dfrac{\delta m_{13}^2}{4E}L}
\r).
\eea
When the condition (\ref{ap-cond-2}) applies,
we find
\bea
P_{\nu_\alpha \to \nu_\alpha} &=&
1
-2 |U_{\alpha 3}^{}|^2 \l(1-|U_{\alpha 3}^{}|^2\r)
-4 |U_{\alpha 1}^{}|^2 |U_{\alpha 2}^{}|^2 
\sin^2
\l(
{\dfrac{\delta m_{12}^2}{4E}L}
\r).
\eqlab{A9}
\eea

\label{app:1}
\setcounter{equation}{0}
\section{Renormalization Group Equations of Yukawa matrices and gauge couplings}
In the MSSM, the RGE of the Yukawa matrices 
in the 1-loop level are \cite{SUSY-RGE}
\bea
\dfrac{d}{dt}\hu & = & \dfrac{1}{\l( 4\pi \r)^2}
     \l[\tr \l( 3 \hu \hu^{\dagger} + \hn \hn^{\dagger} \r) 
      + 3 \hu \hu^{\dagger} + \hd \hd^{\dagger}
     -4\pi\l(\dfrac{16}{3}\alpha_3+3\alpha_2+\dfrac{13}{15} \alpha_1\r) \r]\hu, 
\nn
\\
\dfrac{d}{dt}\hd & = & \dfrac{1}{\l( 4\pi \r)^2}
      \l[\tr \l( 3 \hd \hd^{\dagger} + \he \he^{\dagger} \r)
       + 3 \hd \hd^{\dagger} + \hu \hu^{\dagger}
     -4\pi\l(\dfrac{16}{3}\alpha_3+3\alpha_2+\dfrac{7}{15} \alpha_1\r) \r]\hd, 
\nn
\\
\dfrac{d}{dt}\he & = & \dfrac{1}{\l(4\pi\r)^2}
      \l[\tr \l( 3 \hd \hd^{\dagger} + \he \he^{\dagger} \r)
       + 3 \he \he^{\dagger} + \hn \hn^{\dagger}
     -4\pi\l(3\alpha_2+\dfrac{9}{5} \alpha_1\r) \r]\he.
\nn
\\
\dfrac{d}{dt}\hn & = & \dfrac{1}{\l(4\pi\r)^2}
      \l[\tr \l( 3 \hu \hu^{\dagger} + \hn \hn^{\dagger} \r)
       + 3 \hn \hn^{\dagger} + \he \he^{\dagger}
     -4\pi\l(3\alpha_2+\dfrac{3}{5} \alpha_1\r) \r]\hn,
\eqlab{RGE-high}
\eea
where $t$ is the logarithm of the renormalization scale $\mu$:
\beq
t = \ln \mu .
\eeq
The gauge couplings satisfy the RGE's
\beq
\dfrac{d}{dt}\dfrac{\alpha_i}{\pi} 
= b_i\l(\dfrac{\alpha_i}{\pi}\r)^2,
\eqlab{RGE_a}
\eeq
where $\alpha_i$ are
\beq
\alpha_i = \dfrac{g_i^2}{4\pi}, \mbox{\quad \quad} \l(i=1,2,3 \r),
\eqlab{def_ga}
\eeq
and
\beq
g_1^2 = \dfrac{5}{3}g^{\prime 2}.
\eqlab{def_g1}
\eeq
$g^{\prime}$, $g_2$ and $g_3$ are gauge coupling constant of the
$U(1)_Y$, $SU(2)_L$ and $SU(3)_C$, respectively,
and the coefficient $b_i$ are
\bea
b_1 &=& n_g^{} + \dfrac{3}{20}n_H^{}, \nn \\
b_2 &=& n_g^{} + \dfrac{1}{4}n_H^{} - 3, \nn \\
b_3 &=& n_g^{} - \dfrac{9}{2}.
\eqlab{b_SUSY}
\eea
The factors $n_H^{}$ and $n_g^{}$ are, respectively, the number of 
Higgs doublets and that of fermion generations.
 In this article, we set $n_H^{}=2$ and $n_g=3$.
 Below the right-handed neutrino decoupling scale $(\mu=M_R)$,
the matrix $\hn$ decouples and the RGE of
the effective dimension-five operator, \eqref{RGE_kappa},
takes over.

\label{app:2}
\setcounter{equation}{0}
\section{The phases in the MNS matrix}
We define the eigenvalues of $\kappa$ as
\beq
\kappa_i = |\kappa_i| e^{i \varsigma_i}\,, 
\mbox{\quad \quad}(|\kappa_1|<|\kappa_2|<|\kappa_3| )\,
\eeq
at the $m_Z^{}$ scale.
They are obtained by the unitary transformation
\beq
{\cal U_{\nu}}^T \kappa^{\ast} {\cal U_{\nu}}
 = diag.(\kappa_1^{\ast},\kappa_2^{\ast},\kappa_3^{\ast})\,.
\eeq
On the other hand, the unitary matrix $U_{\nu}$ that
gives real positive neutrino masses obtain
\bea
diag.(m_1,m_2,m_3)
&=& U_{\nu}^T M_{\nu}^{} U_{\nu} \nn \\
&=& U_{\nu}^T {\kappa}^{\ast} U_{\nu} v_{\Su}^2 \nn \\ 
&=& diag.(|\kappa_1|,|\kappa_2|,|\kappa_3|) v_{\Su}^2 \,.
\eea

Hence the matrix $U_{\nu}$ is obtained from ${\cal U_{\nu}}$ by
\beq
U_{\nu} = \cal{U_{\nu}} {\cal P},
\eqlab{c1}
\eeq
where ${\cal P}$ is a diagonal phase matrix.
This phase matrix is obtained as
\bea
U_{\nu}^T \kappa^{\ast} U_{\nu} 
&=& {\cal P}{\cal{U_{\nu}}}^T \kappa^{\ast} \cal{U_{\nu}} {\cal P} \nn \\
&=& {\cal P}
 diag.(\kappa_1^{\ast},\kappa_2^{\ast},\kappa_3^{\ast}) {\cal P} \nn \\
&=& diag.(|\kappa_1|,|\kappa_2|,|\kappa_3|)\,,
\eea
and hence
\beq
{\cal P}  =
\bmaT 
e^{i \frac{\varsigma_1}{2}} & 0 & 0 \\
0 & e^{i \frac{\varsigma_2}{2}} & 0 \\
0 & 0 & e^{i \frac{\varsigma_3}{2}}
\emaT
\,.
\eeq

The MNS matrix is rewritten by using \eqref{c1}
\bea
V_{\rm MNS}^{} &=& U_{\Se}^{\dagger} U_{\nu} \nn \\
            &=& U_{\Se}^{\dagger} \cal{U_{\nu}} {\cal P} \nn \\
            &=& U_{\rm MNS}^{}{\cal P^{\prime}} {\cal P}\,,
\eea
where an additional phase matrix
\beq
{\cal P^{\prime}}  =
\bmaT 
1 & 0 & 0 \\
0 & e^{i \varphi_2^{\prime}}& 0 \\
0 & 0 & e^{i \varphi_3^{\prime}}
\emaT
\,,
\eeq
makes the $U_{e2}$ and $U_{\mu 3}$ elements
real and non-negative
in our phase convention for the MNS matrix.
The Majorana phases of the MNS matrix, $\varphi_2$ and $\varphi_3$,
are now obtained as:
\bea
{\varphi_2}
&=& {\varphi_2^{\prime}}+\dfrac{\varsigma_2-\varsigma_1}{2}\,, \nn \\
{\varphi_3}
&=& {\varphi_3^{\prime}}+\dfrac{\varsigma_3-\varsigma_1}{2}\,. \nn \\
\eea
 These phases are observable, and hence are independent of phase
convention, in lepton-number violating processes.

\end{document}